\documentclass[11pt]{article}
\usepackage{authblk}
\usepackage{hyperref}
\usepackage{natbib}
\usepackage{graphicx}
\usepackage{verbatim}
\usepackage{placeins}

\usepackage[vscale=0.88,hscale=0.8,includehead,vmarginratio=1:9]{geometry}
\usepackage{times}

\newcommand{\erf}[1]{{\mathrm{erf}#1}}
\newcommand{\inverf}[1]{{\mathrm{erf}^{-1}#1}}

\begin{document}
	\title{Calibrating GONG Magnetograms with End-to-end Instrument Simulation II: Theory of Calibration}
	\author[1]{Joseph Plowman}
	\author[2]{Thomas Berger}
	\affil[1]{National Solar Observatory, \url{jplowman@nso.edu}} 
	\affil[2]{University of Colorado at Boulder}
	\date{}
	\maketitle
	\begin{abstract}
		This is the second of three papers describing an `absolute' calibration of the GONG magnetograph using an end-to-end simulation of its measurement process. In the first paper, we described the GONG instrument and our `end-to-end' simulation of its measurement process. In this paper, we consider the theory of calibration, and magnetograph comparison in general, identifying some of the significant issues and pitfalls. The calibration of a magnetograph is a function of whether or not it preserves flux, independent of its spatial resolution. However, we find that the one-dimensional comparison methods most often used for magnetograph calibration and comparison will show dramatic differences between two magnetograms with differing spatial resolution, {\em even if they both preserve flux}. Some of the apparent disagreement between magnetograms found in the literature are likely a result of these instrumental resolution differences rather than any intrinsic calibration differences. To avoid them, spatial resolution must be carefully matched prior to comparing magnetograms or making calibration curves. In the third paper, we apply the lessons learned here to absolute calibration of GONG using our `end-to-end' measurement simulation.
	\end{abstract}

	\section{Introduction \& Calibration Requirement}\label{sec:intro}

	A variety of instruments currently observe the solar magnetic field, and there have been numerous comparisons, both between magnetographs and with {\em in situ} measurements of the magnetic field. \cite{Riley_comparison2014}, for example, compare seven solar magntographs and find that their measurements (called magnetograms) can differ by an order of magnitude. They find HMI, VSM, and GONG can differ by factors of $\sim 2$. \cite{VirtanenMursula2017}, on the other hand, compare observations six of those same observatories (sans GONG) in terms of a harmonic expansion, and find better agreement, at least between SDO/HMI and SOLIS/VSM -- that relationship is reasonably approximated by a factor of 0.8. However, \cite{LinkerOpenFlux2017} compares the open flux in magnetic field extrapolations (produced from magnetograms) to in situ measurements and finds a discrepancy of 2.0 or more across all instruments. This suggests that the magnetographs may be systematically underestimating their measurements. Both \cite{LinkerOpenFlux2017} and \cite{Riley_comparison2014} provide a number of references documenting this practice. The variability of these comparison results, and the importance of the magnetograms for space weather, call for a better calibration of the magnetograms. However, an `absolute' calibration has not been done so far owing to a lack of {\em in situ} `ground truth' observations corresponding to the magnetograph measurements -- e.g., magnetometer measurements {\em in} the photosphere.

	The work described in this paper is part of a project to create an `absolute' calibration for a solar magnetograph using a solar 3D MHD simulation and an `end-to-end' simulation of the measurement process. The calibration can be `absolute' because we possess, in the 3D MHD simulation, the `ground truth' corresponding to the simulated measurements. The magnetograph in question is the National Solar Observatory (NSO) Global Oscillations Network Group \citep[GONG;][]{HarveyEtal_GONG1996} instrument and the 3D MHD simulation is a MURaM \citep[MURaM;][]{Rempel2015} simulation of a sunspot and its environs. We have described the MURaM simulation and the GONG simulation already in \cite{PlowmanEtal_2019I}, as well as giving a detailed introduction to the problem. In the current paper \citep{PlowmanEtal_2019II}, we describe in detail how to use the MURaM simulation and the GONG simulation to construct a calibration, and some of the issues surrounding construction of such calibrations, and magnetogram comparison in general. In the process, we will shed light on both the inconsistent results of existing comparisons and genuine source of `miscalibration' of magnetographs in general, compared with the `ground truth' fluxes.

	The issues with magnetograph calibration arise most frequently in the context of the global field extrapolations used in space weather prediction: almost all papers discussing the issue (including those cited so far) make it clear that this topic is their primary concern. This is especially true for GONG magnetograms: of the papers on NASA ADS citing the primary GONG instrument paper \citep[GONG;][]{HarveyEtal_GONG1996} which use the GONG magnetograms, over 75\% use them for global field extrapolations. One of the most important and representative of these is the WSA-ENLIL+Cone modeling used by the Community Coordinated Modeling Center (CCMC) located at NASA Goddard space flight center \citep{MaysEtal_SoPh2015}. It is based on a potential field source surface (PFSS) extrapolation \citep[in the WSA component -- see][]{ArgePizzo_JGR2000}, as are most other extrapolations used in space weather.

	The quantities measured by longitudinal magnetographs are most often referred to as fields. However, they are related to the magnetic fields via a {\em signed} average (i.e., an integral) of the line-of-sight component of the field over a highly flattened volume on the surface of the sun (the photosphere); for GONG pixels, these regions are $\sim 150$ km deep and $\sim 6000$ km across, so the volume integral is nearly a surface integral. They are therefore most analogous to fluxes, which are integrals over a surface of the component of the magnetic field perpendicular to that surface (for vertical viewing angles, this surface normal component is almost identical to the line-of-sight component, making the measurements even more similar to fluxes). 

	Moreover, in Appendix \ref{app:PFSS} we show that a calibrated magnetogram will give the same PFSS extrapolation as its real ground truth so long as it has the same flux at larger spatial scales. A `calibration' which does not have the same flux at larger spatial scales, on the other hand, will give an incorrect extrapolation {\em even if the real field is potential}. Although the PFSS may be overly simplistic for most solar cases, it is still important because it is the basis for most space weather models in use today. Not only that, a calibration should be general, and give valid results for simple test cases as well as complex ones. Calibrations which do not preserve fluxes can therefore be rejected on these grounds alone.
		
	We therefore require that a calibrated magnetogram does not add, remove, amplify, or attenuate flux compared to the ground truth. It can move flux around (e.g., due to a PSF), and those fluxes may cancel with an opposing flux moved in from a neighboring region, but cannot remove or add flux {\em ex nihilo}; it has to go to or come from somewhere. We tend to use the term `flux' to refer to the measurements as a reminder of this requirement, but sometimes fall back on the more conventional term `field' instead. Whichever term is used, it should be remembered that the magnetograph measurements are always an average of the line-of-sight component of the solar magnetic field over a (close to two dimensional) surface in the photosphere. Units reported are typically Gauss, but this should not distract from the fact that they are integrated quantities; the length units resulting from the volume integration have effectively been normalized out.


	\section{Theory of Calibration}\label{sec:calibrationtheory}
	
\begin{figure}
		\begin{center}\includegraphics[width=\textwidth]{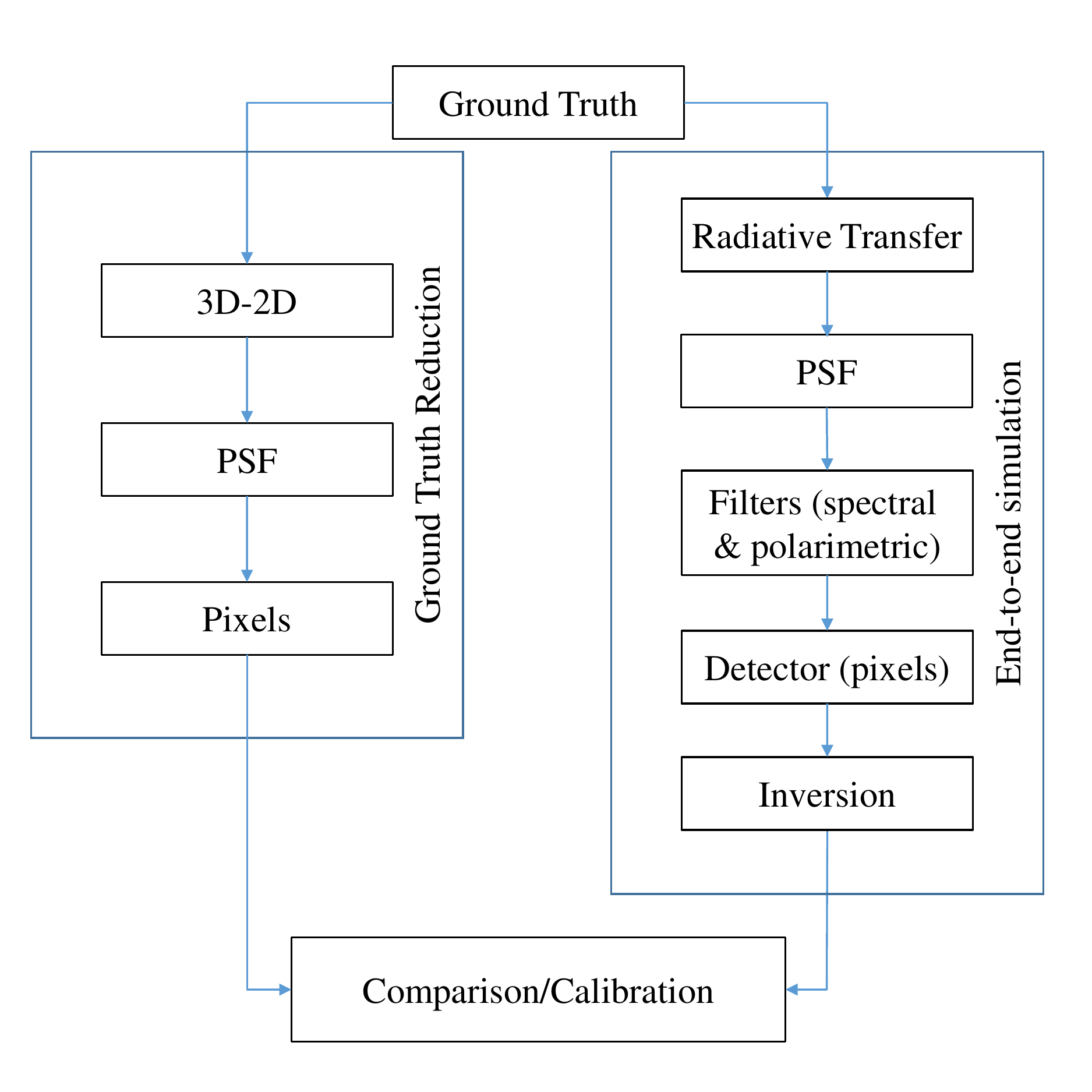}\end{center}
		\caption{Flow chart of the calibration process. On the right is the end-to-end simulation of the measurement process, while on the left is the `ground truth reduction' which places the ground truth in a form that can be directly compared with the measurements. This is necessary because some parts of the measurement process are irreversible, as described in the text.}\label{fig:calibration_boxdiagram}
	\end{figure}

	In \cite{PlowmanEtal_2019I} we produced a set of `synthetic' GONG flux (i.e., average line-of-sight field or average flux density)  measurements for which we possess the `ground truth' fluxes (or magnetic fields -- i.e., the magnetic fields from the MURaM snapshot which was the input to the instrument simulator). We want to use this information to relate measurements to the `correct' value based on the ground truth. Here we encounter an issue:

	The synthetic flux measurements have been produced by our simulation of the measurement process, represented in a high-level schematic form by the right-hand side of Figure \ref{fig:calibration_boxdiagram}. This measurement process simulation is `end-to-end', in the sense that it proceeds all the way from one end (the sun) to the other (the measurements) and simulates each step of the measurement process. This is possible because every step of this forward process is well defined by the physics and its output is uniquely determined by the inputs (the outputs of the previous step). The reverse is not true because many of these `forward' steps are inherently irreversible: their inputs are not uniquely determined by their outputs: information is therefore lost in the forward steps. 

	If the amount of information lost is small, we may be able to produce a proxy of the ground truth by replacing the lost information with post facto constraints such as smoothness. However, such processes are likely to amplify the noise in unexpected fashions, and post facto constraints are not a good substitute for the lost information. Deconvolving an instrument PSF is an illustrative example of this. That problem is not ill-posed {\em in principal} because there are as many deconvolved pixels as there are data points, so the forward matrix is square. However, it is ill-posed {\em in practice} because adjacent pixels have very similar spatial responses, which makes the forward matrix nearly singular; it is therefore {\em ill-conditioned}. As a result, any noise in the data produces rapid spatially correlated oscillations in the deconvolution. Thus, the results of the deconvolution are not well-defined by the physics, and the outputs are not well determined by the inputs in the way that the forward problem is. The only way to remove the oscillations is to apply additional constraints such as smoothness (but this can also make the deconvolution nonlinear): the noise and PSF together have effectively effaced the high-frequency information.

	Consequently, we cannot relate the measurements back to the ground truth in the same systematic, end-to-end fashion; `End-to-end calibration' can only mean `calibration based on end-to-end simulation of the measurement process'. What we can do instead is relate the ground truth forward in a way which preserves the features which are most relevant for the intended uses of the data. In this case, we have argued, those features are the fluxes. We call this process `ground truth reduction'. 

	Irreversibility primarily manifests itself in that there is not just one ground truth flux value that contributes to a given synthetic measurement, but many of them. The ground truth is a high-resolution 3-dimensional array of magnetic flux values, while the observations are a low-resolution 2-dimensional array of magnetic flux measurements. For purposes of exposition, this degeneracy can therefore be split into two categories:

	\begin{enumerate}
		\item Plane-of-sky degeneracy owing to the drastically higher resolution of the MURaM snapshot compared with GONG.\label{enum:posdegen}
		\item Line-of-sight degeneracy owing to the range of depths sampled by the radiative transfer.\label{enum:losdegen}
	\end{enumerate}

	Ground truth reduction resolves these degeneracies, by producing a `ground truth magnetogram' which can be put on a one-to-one correspondence with the synthetic GONG magnetogram. In essence, it should represent, not the magnetogram that would be produced if the instrument were `perfect', but the one produced if it were `perfectly calibrated'.

	This process is shown in Figure \ref{fig:calibration_boxdiagram}. In the comparison stage, we investigate the relationship between the measurements and the ground truth by way of a two-dimensional pixel-to-pixel comparison: One dimension is the magnetic fluxes from the ground truth and the other is the corresponding fluxes from the synthetic magnetogram. Graphically, such comparisons can be represented by scatter plots; a number of examples can be found in subsequent sections.

	From the scatter plots, we fit calibration curves which are one dimensional functions of the measured flux values, such that for a measured flux (at pixel location $\{x,y\}$) $\phi_\mathrm{meas}(x,y)$, the calibrated flux, $\phi_\mathrm{cal}(x,y)$, is given by
	\begin{equation}
		\phi_\mathrm{cal}(x,y) = f_\mathrm{cal} (\phi_\mathrm{meas}(x,y)).
	\end{equation}
	To account for differences between active regions and the rest of the sun, we will compute separate curves for each of those categories, and for each inclination simulated. To apply this calibration to a magnetogram, we simply evaluate these functions for the measurement at every pixel in the magnetogram, interpolating between inclination angles. This will be fully implemented, and demonstrated, in the final paper of the series \citep{PlowmanEtal_2019III}.

	A higher-dimensional, `spatially aware', calibration approach, where each pixel's calibrated flux depends on the measurements of the surrounding pixels, would likely be capable of producing a better calibration than this method. However, it would be much more involved to produce and validate. A neural network, for example, would require a large number of simulations to train (we only have one or two available) and even then there is an increased danger that the results are biased by subtle details of the simulation that may not be physical: The network would `restore' the information lost as part of the measurement process by a kind of `statistical interpolation', substituting details of the 3D MHD simulation (e.g., MURaM).

	\subsection{Spatial resolution effects: From high resolution ground truth to low resolution}\label{sec:spatialresolution}

	Let us assume for now that the ground truth has been reduced to the two plane-of-sky dimensions \citep[we will cover that part of the reduction in the final paper of the series,][]{PlowmanEtal_2019III}, but is at higher resolution than the measurements. The first question, then, is how to produce from this high-resolution ground truth magnetogram a low-resolution version where each pixel is on a one-to-one correspondence with the synthetic magnetogram. One obvious suggestion is to resample the ground truth magnetogram to the pixel grid of the synthetic magnetogram. However, we find that an issue arises if the ground truth is not also convolved with the instrument PSF, owing to the resulting resolution difference.
	
\begin{figure}
		\begin{center}\includegraphics[width=\textwidth]{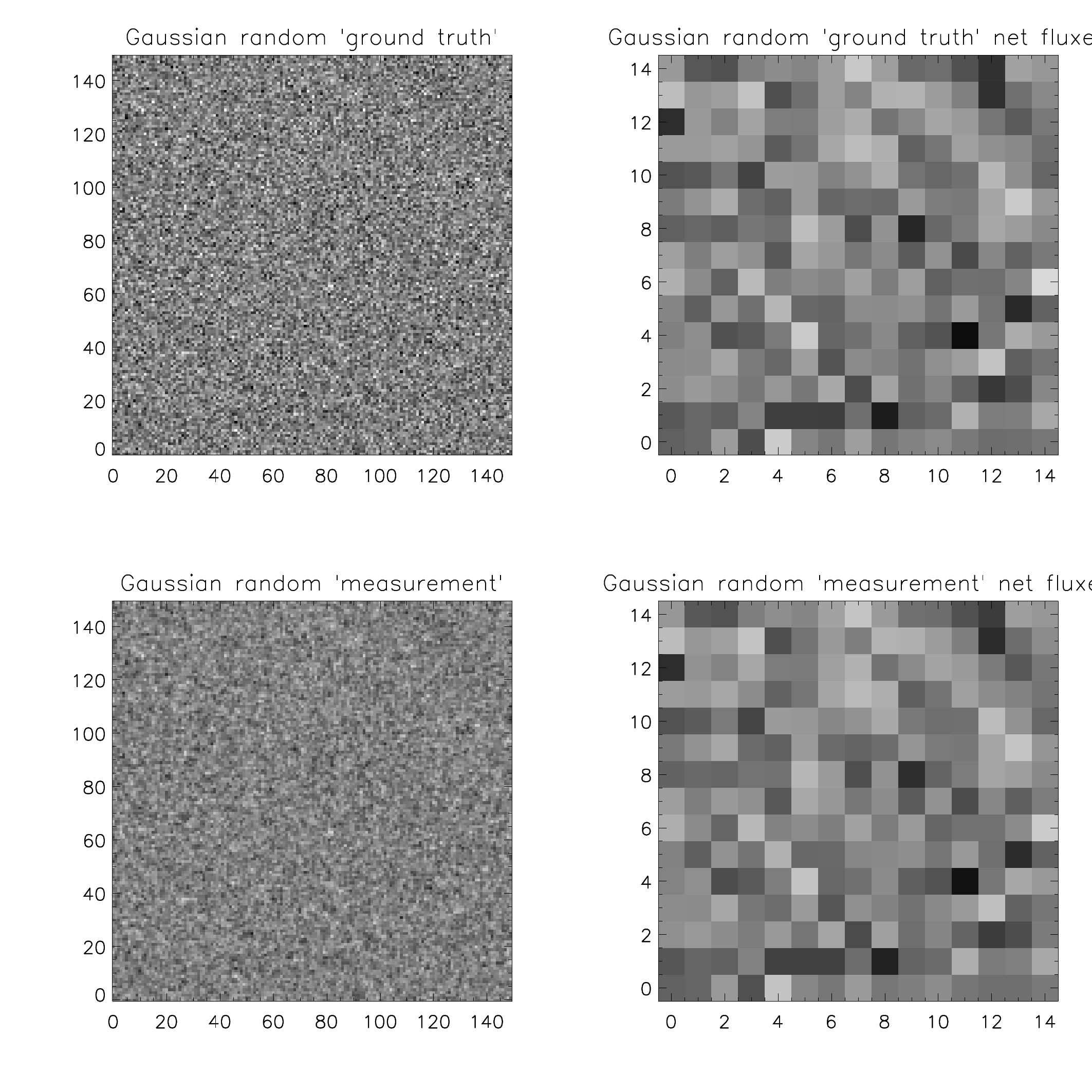}\end{center}
		\caption{Simple ground truth and measured `magnetogram' examples used to introduce PSF issues in Section \ref{sec:spatialresolution}. Left panels: ground truth (top), which is random and uncorrelated from pixel to pixel, and `measurement' (bottom), which differs from ground truth only by a linear PSF convolution. Right panels: net fluxes of each $10\times 10$ pixel region in ground truth (top) and measurement (bottom), obtained by summing over each region. Notice that spatial resolution and the `dynamic range' of the measurement are both lower than the ground truth in the left panels (brightest pixels are not as bright, darkest pixels are not as dark, etc), while in the right panels they are identical -- See Figures \ref{fig:simplenetfluxexample_scatterplot} and \ref{fig:simpleexample_scatterplot} for scatter plots. }\label{fig:groundtruthpsf_perpixel}
	\end{figure}

\begin{figure}
		\begin{center}\includegraphics[width=0.75\textwidth]{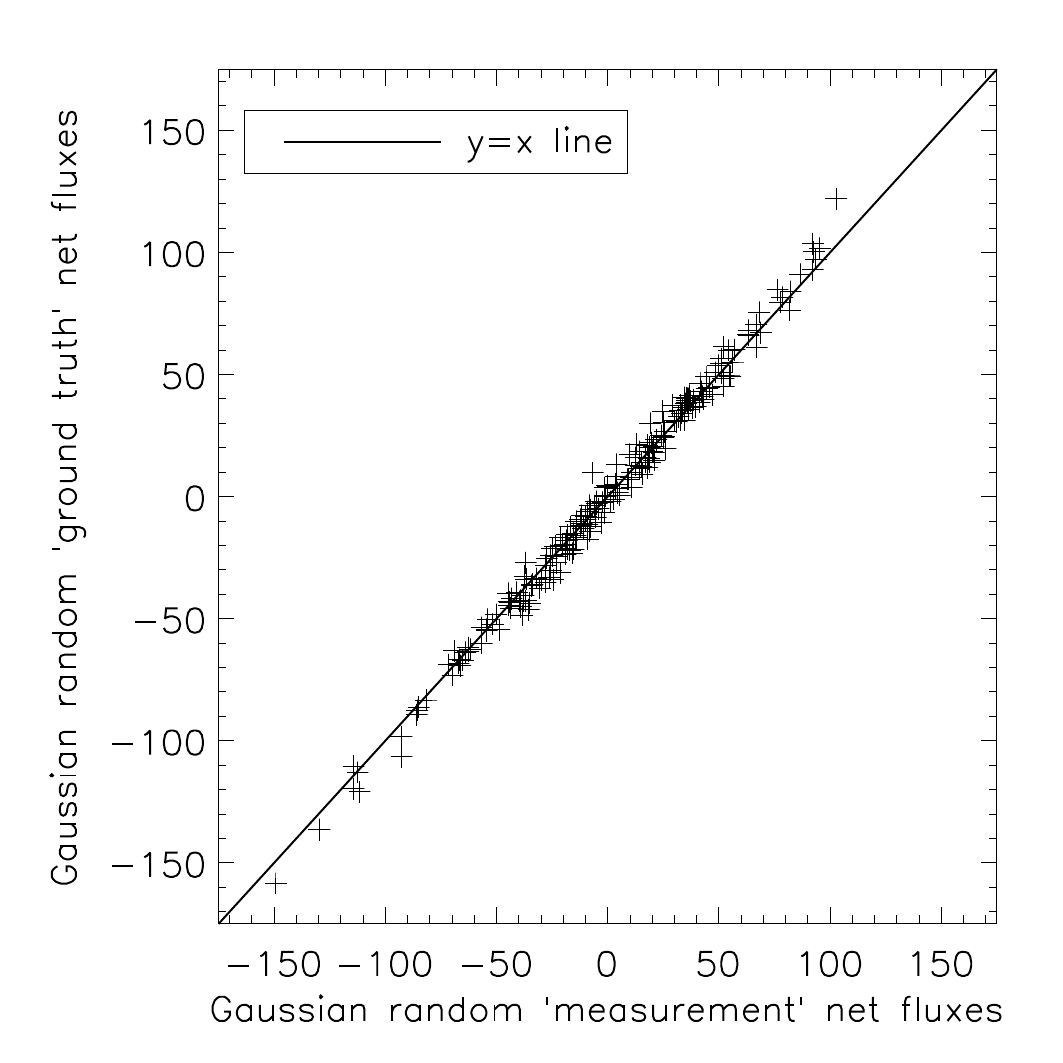}\end{center}
		\caption{Scatterplot of ground truth vs. `measured' large-scale fluxes for example shown in right panels of Figure \ref{fig:groundtruthpsf_perpixel}. Measured differs from ground truth in this example only in application of a linear, normalized PSF. The scatter plot shows that the large-scale fluxes are not affected by the PSF. The per-pixel fluxes, however, are -- see scatterplot in Figure \ref{fig:simpleexample_scatterplot}.}\label{fig:simplenetfluxexample_scatterplot}
	\end{figure}

	This section investigates the question in considerable detail, but first we illustrate that it occurs even when the measurement is `perfect' except for having a PSF: the {\em only} difference between the ground truth and measurement is the PSF. The issue arises is due to lack complete correlation at the pixel scale of the instrument, so a completely uncorrelated case (drawn from a Gaussian random distribution) is used for this illustration (Section \ref{sec:linearmeasurement_solarexamples} shows additional solar examples). The left panels of Figure \ref{fig:groundtruthpsf_perpixel} show an example of this `ground truth' and `measured' magnetogram. For this example, the PSF is a two-dimensional Gaussian with standard deviation 0.5 pixels applied to it. It is normalized to one and applied by a linear convolution, so it does not add or remove flux, only spreads it out. The right panels of Figure \ref{fig:groundtruthpsf_perpixel} show the large-scale ($10\times 10$ pixel regions) flux, obtained directly from the ground truth and measurement in the left panels by summing over each region; no other operation (and in particular no calibration curve or factor of any kind) has been applied. The large-scale flux for the ground truth and measured magnetograms clearly match: most compellingly, the scatter plot (Figure \ref{fig:simplenetfluxexample_scatterplot}) shows a clear, linear relationship with a slope of two and no offset. Thus, there is no need of any `calibration factor' to restore the correct net flux in this case. 

\begin{figure}
		\begin{center}\includegraphics[width=0.75\textwidth]{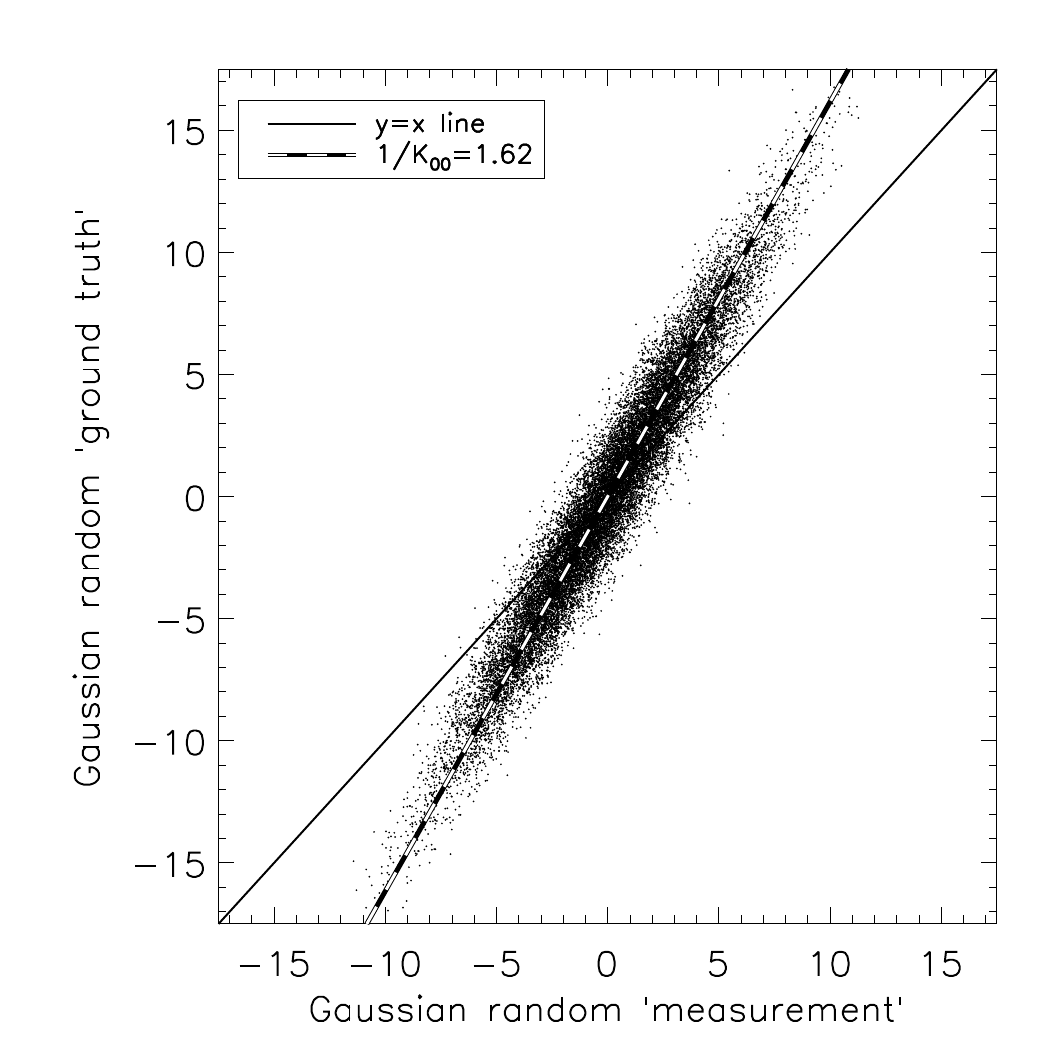}\end{center}
		\caption{Scatterplot of ground truth vs. `measured' per-pixel fluxes for example shown in left panels of Figure \ref{fig:groundtruthpsf_perpixel}. Measured differs from ground truth in this example only in application of a linear, normalized PSF. This per-pixel flux scatter plot is affected by the spatial resolution difference between measurement and ground truth, unlike the large-scale fluxes (Figure \ref{fig:simplenetfluxexample_scatterplot}). This scatterplot appears to imply a calibration factor of $\sim 1.6$, but application of that factor to the measurements results in a bad large-scale flux calibration (see Figure \ref{fig:groundtruthpsf_badcal_netflux}).}\label{fig:simpleexample_scatterplot}
	\end{figure}

\begin{figure}
		\begin{center}\includegraphics[width=0.75\textwidth]{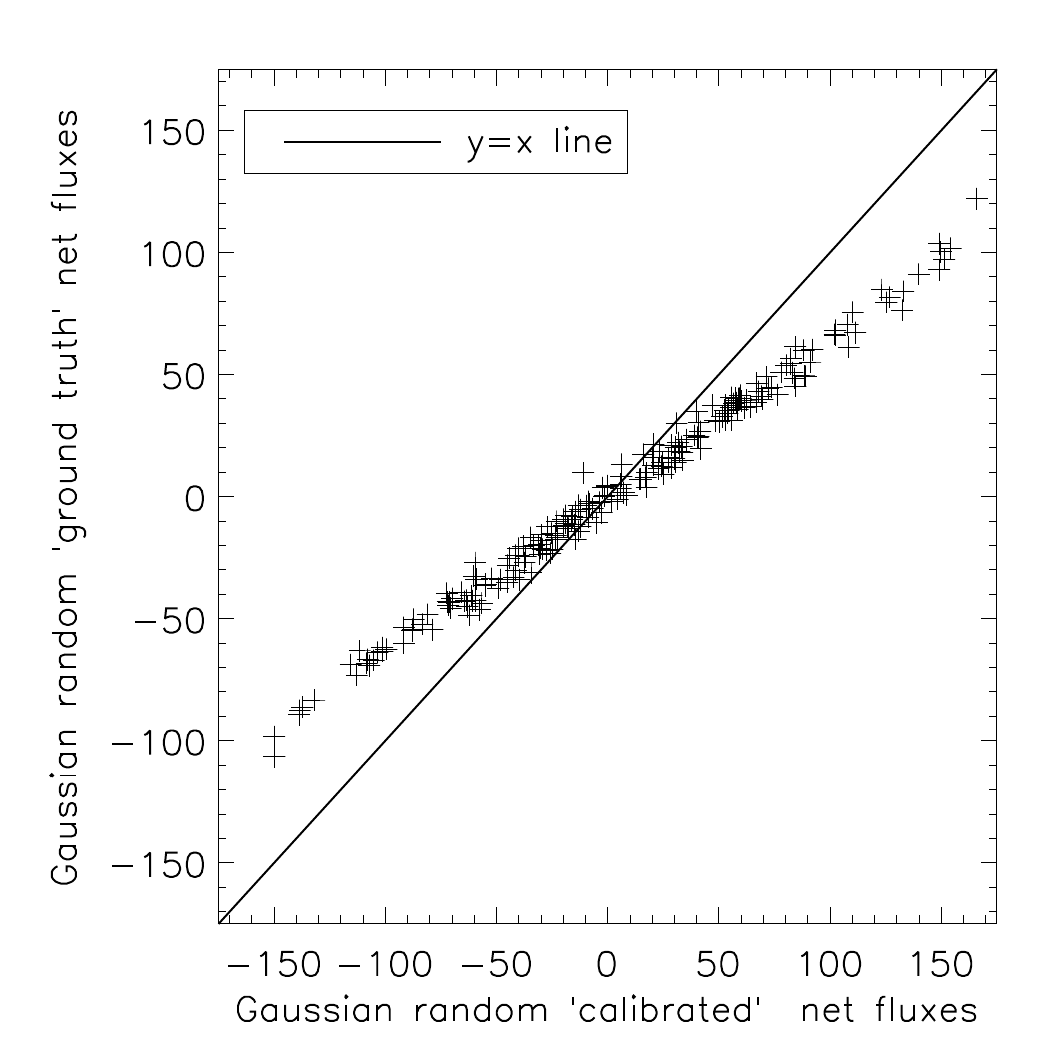}\end{center}
		\caption{Bad net flux calibration resulting from application factor suggested by per-pixel scatterplot (Figure \ref{fig:simpleexample_scatterplot}) to measurements of Figure \ref{fig:groundtruthpsf_perpixel}. The net fluxes are now too large by a factor of $\sim 1.6$, illustrating the issue with attempting to create a calibration curve from ground truth whose spatial resolution does not match that of the measurements.}\label{fig:groundtruthpsf_badcal_netflux}
	\end{figure}

	Figure \ref{fig:simpleexample_scatterplot} shows what happens when we plot the ground truth and measurement directly against each other as a scatter plot, without matching the spatial resolutions or binning to large scale: the scatter plot is very different from that for the large-scale fluxes, showing a clear slope of $\sim 1.6$, {\em not} the factor of 1 required to for the large-scale fluxes to match. If we take this as the `calibration' factor, apply it to the measurements, and recompute the large-scale net fluxes from the resulting `calibrated' measurements, they are now {\em too high} by a factor of $\sim 1.6$ (this factor is related to the details and size of the PSF; see subsequent sections). This is shown by Figure \ref{fig:groundtruthpsf_badcal_netflux}. 

	Thus, when spatial resolution of the `ground truth magnetogram' does not match that of the `synthetic magnetogram', the shape of the point cloud does not, in general, reflect conservation (or nonconservation) of the fluxes in the magnetograms. A `calibration' curve made from such a point cloud will amplify the fluxes, as Figure \ref{fig:groundtruthpsf_badcal_netflux} has shown. This is unacceptable since preservation of flux is the primary calibration goal. The solution is to resample the ground truth magnetogram to the same resolution as the synthetic magnetogram.

	\subsubsection{Linear measurement model}\label{sec:largescalelinear_fluxconservation}

	To demonstrate that these PSF issues are general and not due to pathologies of a particular instrument, spectral line, or other factors, we first use an example case with instrumental effects that are as benign as possible. We call this example the `linear measurement model', which we define as follows: 
	
	Denoting the original `ground truth' or `actual' fluxes by $\phi^0_{ij}$, and the corresponding `measured' fluxes by $\phi^m_{ij}$, the $\phi^m_{ij}$ are related to the $\phi^0_{ij}$ only by a constant `miscalibration' factor, $c_m$, a convolution (representing the PSF), a measurement flux bias $\phi^m_0$, and a random measurement error $\Delta \phi^m_{ij}$ which has mean 0:

	\begin{equation}\label{eq:linearmeasurement_model}
		\phi^m_{ij} = \phi^m_0+\Delta \phi^m_{ij} + c_m \sum_{kl}w_{ij,kl}\phi^0_{ij}.
	\end{equation}
	$w_{ij,kl}$ expresses the convolution as a full-dimensional linear operator. It is related to the convolution kernel, $K_{mn}$, by $w_{ij,kl} = K_{k-i,l-j}$ (We have defined the kernel indices as running from $-n$ to $n$, for a kernel size of $2n-1$ and a kernel center at $m=0$, $n=0$). Note in particular that $w_{ij,ij} = K_{00}$ is the PSF weight of the central pixel, and that $\phi^m_{ij}$ has lower resolution than $\phi^0_{ij}$ due to the PSF, both of which take on a central role in the ensuing discussion. 
	
	To calibrate the data, we seek a calibration curve, $f_\mathrm{cal}(x)$, which is a one-dimensional function of the measured fluxes such that the `calibrated' fluxes $\phi^c_{ij}$ are 
	\begin{equation}
		\phi^c_{ij} = f_\mathrm{cal}(\phi^m_{ij}).
	\end{equation}

	We first make the very important point that, due to the normalization of the PSF, for any region large enough for edge effects to be small (specifically, the {\em changes} in values at the edge of the region caused by the PSF must be small compared to the region's net flux), we have
	\begin{equation}\label{eq:PSF_fluxconservation}
		\sum_{ij\ \mathrm{region}}\sum_{kl}w_{ij,kl} \phi^0_{ij} = \sum_{ij\ \mathrm{region}} \phi^0_{ij} \equiv \Phi_\mathrm{region}.
	\end{equation}
	In other words, the PSF has {\em no effect} on the net flux over large enough (compared to the PSF) regions; it only changes the pixel-scale fluxes near the edges of the regions. 	The measured flux over the region is then
	\begin{equation}
		\sum_{ij\ \mathrm{region}} \phi^m_{ij} = \sum_{ij\ \mathrm{region}} (\phi^m_0+\Delta \phi^m_{ij}) + c_m\Phi_\mathrm{region}.
	\end{equation}
	Therefore,
	\begin{equation}\label{eq:fluxconservation_general}
		\Phi_\mathrm{region} = \sum_{ij\ \mathrm{region}}\frac{\phi^m_{ij}-\phi^m_0}{c_m}
	\end{equation}
	minus the random measurement error term ($\Delta \phi^m_{ij}/c_m$). 

	Now, consider a generic calibration curve, where the calibrated value at each pixel, $\phi^c_{ij}$, is a function $f_\mathrm{cal}$ only of the measured value at that pixel ($\phi^m_{ij}$) and precomputed constants -- without loss of generality, this can be represented as a Taylor series:
	\begin{equation}
		f_\mathrm{cal}(\phi^m_{ij}) = \sum_{n=0}^\infty q_n \big(\phi^m_{ij}\big)^n.
	\end{equation}
	The net flux of the region is related to the measured flux by Equation \ref{eq:fluxconservation_general}). Therefore, flux conservation requires that all parts of this calibration except for 
	\begin{equation}
		\frac{\phi^m_{ij}}{c_m}-\frac{\phi^m_0}{c_m}
	\end{equation}
	sum to zero when the calibrated values are summed over any region large enough for edge effects to be small. Therefore, if we want the calibration to conserve large-scale flux for any input ground truth, we must have
	\begin{equation}
		q_0 = -\frac{\phi^m_0}{c_m}, \qquad q_1 = \frac{1}{c_m},
	\end{equation}
	and all other $q_n$ must be zero. In other words, there is a unique calibration curve which conserves net flux for all ground truth magnetograms, and it is
	\begin{equation}\label{eq:linear_fluxconserving_curve}
		f_\mathrm{cal}(\phi^m_{ij}) = \frac{\phi^m_{ij}}{c_m}-\frac{\phi^m_0}{c_m}.
	\end{equation}
	Notice in particular that this has no dependence on the PSF or the spatial resolution of the instrument in general (with the obvious caveat that the spatial resolution must be higher than the large-scale features in question).

	Now, consider finding a calibration curve based on the relationship between the $\phi^m_{ij}$ and the $\phi^0_{ij}$ -- i.e., the former defined by Equation \ref{eq:linearmeasurement_model} and consequently has the PSF applied to it, while the latter does not. This relationship can be seen from Equation \ref{eq:linearmeasurement_model} by splitting the $\phi^0_{ij}$ term in the summation apart from the other terms:

	\begin{equation}\label{eq:mij_grouped}
		\phi^m_{ij} = \phi^m_0+\Delta \phi^m_{ij} + K_{00}c_m \phi^0_{ij} + c_m \sum_{kl\neq ij}w_{ij,kl}\phi^0_{kl},
	\end{equation}
	where $\sum_{kl\neq ij}$ means that the term with $k=i$ and $l=j$ is omitted from the summation. This allows us to match the $\phi^m_{ij}$ on the left-hand side with the $\phi^0_{ij}$ on the right-hand side. It matches how the `measured' and `actual' values are plotted against each other in a scatter plot: for each pixel indexed by $i$ and $j$, we plot $\phi^0_{ij}$ against $\phi^m_{ij}$.
	
	The results of this procedure will in general depend on the ground truth values, $\phi^0_{ij}$  used to construct the calibration. To explore this, we consider an illustrative test case that is amenable to analytical treatment.  At GONG's resolution, much of the sun resembles `salt and pepper' noise, uncorrelated between adjacent pixels: so, consider 
	\begin{equation}\label{eq:randomaij}
		\phi^0_{ij} = \phi^0_0+r_{ij},
	\end{equation}
	where the $r_{ij}$ are a set of random values which are completely uncorrelated from each other and have median 0. Remembering that the PSF is normalized so that $\sum_{ij} w_{ij,kl}=1$, Equation \ref{eq:mij_grouped} becomes

	\begin{equation}
		\phi^m_{ij} = c_m \phi^0_0 + c_m r_{ij}K_{00} + c_m \sum_{kl\neq ij}w_{ij,kl}r_{kl} + \phi^m_0 + \Delta \phi^m_{ij}.
	\end{equation}
	The $r_{kl}$ are uncorrelated with each other, so each $\sum_{kl\neq ij}w_{ij,kl}r_{kl}$ is completely uncorrelated with $r_{ij}$: the $ij$th pixel's $r_{ij}$ has been specifically removed from the sum. Although it is true that adjacent $\phi^m_{ij}$ will tend to be correlated with each other due to the PSF, the point clouds do not plot $\phi^m_{ij}$ vs. adjacent $\phi^m_{ij}$. Instead, they plot $\phi^m_{ij}$ vs $\phi^0_{ij}$, and adjacent $\phi^0_{ij}$ are completely uncorrelated with each other in this example case: while adjacent pixels will tend to be close together in the $\phi^m_{ij}$ axis of the scatter plots (the x axis, in this work), their positions in the $\phi^0_{ij}$ axis of the scatter plots (the y axis, in this work) will be completely random due to the lack of correlation between the $\phi^0_{ij}$. This adjacency therefore doesn't contribute to any trend: it is only the $\phi^m_{ij}$ vs. $\phi^0_{ij}$ relationship that provides the correlation, and therefore the calibration curve extracted from the point cloud. And since $r_{kl}$ and $\Delta \phi^m_{ij}$ are both randomly distributed with median zero, they contribue nothing to the calibration curve on average. We can therefore read off the $\phi^m_{ij}$ vs. $\phi^0_{ij}$ relationship as

	\begin{equation}
		\phi^m_{ij} = c_m(\phi^0_0-\phi^0_0 K_{00})+\phi^m_0 + c_m K_{00} \phi^0_{ij}.
	\end{equation}
	In other words, the slope is $c_mK_{00}$ and the offset is  $c_m(\phi^0_0-\phi^0_0 K_{00})+\phi^m_0$. We can use this to compute `calibrated' fluxes, $\phi^c_{ij}$, from the $\phi^m_{ij}$, according to
	
	\begin{equation}\label{eq:naivecal_noactualpsf}
		f_\mathrm{cal}(\phi^m_{ij}) = \phi^m_{ij}\frac{1}{c_m K_{00}} - \frac{\phi^m_0 + c_m(\phi^0_0-\phi^0_0 K_{00})}{c_m K_{00}}.
	\end{equation}
	
	Notice that we do {\em not} obtain the unique general flux-conserving calibration curve of Equation \ref{eq:linear_fluxconserving_curve}: The slope has an extra factor of $1/K_{00}$ and the offset is different as well. The scatter plot shown in Figures \ref{fig:simpleexample_scatterplot} is of this type. In that case, $K_{00} \approx 0.6$, $c_m=1$, and $\phi^m_0=\phi^0_0=1$, and the per-pixel scatter plot made from plotting $\phi^m_{ij}$ vs. $\phi^0_{ij}$ exactly matches the slope given by Equation \ref{eq:naivecal_noactualpsf} ($1/K_{00}\approx 1.6$). Attempting to use this point cloud to calibrate measurements will result in an erroneous large-scale flux, as Figure \ref{fig:groundtruthpsf_badcal_netflux} demonstrates.
	
	The curve guaranteeing large-scale flux conservation (Equation \ref{eq:linear_fluxconserving_curve}), on the other hand, has a slope of unity and no offset, as it should: the PSF has been explicitly normalized to one (so it only moves flux around, rather than adding or removing it) and there is no miscalibration ($c_m=1$), so the large-scale flux of the `measurement' and `ground truth' match in this case. This is exactly what was shown in Figure \ref{fig:simplenetfluxexample_scatterplot}.
	
	Therefore, even if the measurement process is linear, this attempted calibration only works (i.e., the `calibrated' net flux of any given subregion is equal to the real net flux of that subregion), if either
	\begin{itemize}
		\item The average flux of each subregion being calibrated exactly matches the average flux, $\phi^0_0$, of the {\em entire} ground truth ($\Phi'_\mathrm{net}/n'_\mathrm{tot}=\phi^0_0$), or
		\item $K_{00}=1$, in which case the PSF must be a delta function (all other parts of the kernel are zero due to normalization).
	\end{itemize}

	This issue is a feature of the pixel-to-pixel comparison method, and not an artifact of fitting procedure. The slope shown in the scatter plots is clear, and we have not had to resort to any fitting method to demonstrate the correlation mathematically (although we do consider two specific fitting procedures in Appendix \ref{app:fittingmethods}). Its source is the spatial resolution mismatch between the ground truth (with no PSF applied to it) and the measurements: 
	
	Simply put, the pixel-to-pixel comparison looks for correlation between $\phi^m_{ij}$ and $\phi^0_{ij}$, and if the $\phi^0_{ij}$ are dominated by random variation (there is {\em no} correlation between $\phi^0_{ij}$ and its neighbors) there is only one term in the $\phi^m_{ij}$ (i.e., the right-hand side of Equation \ref{eq:linearmeasurement_model}) which has a correlation with $\phi^0_{ij}$ -- the term with $kl = ij$, and its coefficient is $1/(c_m K_{00})$. The correlation coefficient from this comparison method is therefore $1/(c_m K_{00})$, {\em not} $1/c_m$ as required by flux conservation (Equation \ref{eq:linear_fluxconserving_curve}. 
	
	We now show that the solution is to convolve the ground truth with a PSF so that its spatial resolution matches that of the synthetic measurements. That is, we convolve $\phi^0_{ij}$ with the PSF before comparing with $\phi^m_{ij}$, producing, the `reduced' ground truth flux $\phi^r_{ij}$:
	\begin{equation}\label{eq:resmatch_linearcomparison}
		\phi^r_{ij} = \sum_{kl}w_{ij,kl}\phi^0_{kl}.
	\end{equation}
	The model values are still
	\begin{equation}
		\phi^m_{ij} = \phi^m_0+\Delta \phi^m_{ij} + c_m \sum_{kl}w_{ij,kl}\phi^0_{kl} = \phi^m_0 + \Delta \phi^m_{ij} +c_m\phi^r_{ij},
	\end{equation}
	So now, the relationship between measurement ($\phi^m_{ij}$) and ground truth ($\phi^r_{ij}$) is linear with a slope of $c_m$ and an offset of $\phi^m_0$ (recall that $\Delta \phi^m_{ij}$ are uncorrelated random measurement errors); These values can be trivially read off a scatter plot. To obtain the calibrated values, we simply subtract the offset and divide by the slope:
	\begin{equation}\label{eq:resmatch_calcurve}
		\phi^{c}_{ij} = \frac{\phi^m_{ij}-\phi^m_0}{c_m} = \frac{\phi^m_{ij}}{c_m}-\frac{\phi^m_0}{c_m}.
	\end{equation}
	This is exactly the calibration curve which conserves flux for all ground truth found in Equation \ref{eq:linear_fluxconserving_curve}. Unlike before, there is no dependence on the ground truth fluxes ($\phi^0_{ij}$) used to produce the calibration, and the curve works for any other ground truth, $\phi^{0'}_{ij}$:

	\begin{equation}
		\phi^{c'}_{ij} = \frac{\phi^{m'}_{ij}-\phi^m_0}{c_m} = \frac{1}{c_m}(c_m\sum_{kl}w_{ij,kl}\phi^{0'}_{kl}+\Delta \phi^m_{ij}) = \sum_{kl}w_{ij,kl}\phi^{0'}_{kl}+\frac{\Delta \phi^{m'}_{ij}}{c_m} =\phi^{r'}_{ij}+\frac{\Delta \phi^{m'}_{ij}}{c_m}.
	\end{equation}
	In either case, this produces the correct net flux for any region large enough for edge effects to be small (defined as before): neglecting the noise term, which only scales with $\sqrt{n_\mathrm{region}}$,
	\begin{equation}
		\sum_{ij\ \mathrm{region}} \phi^c_{ij} =  \sum_{ij\ \mathrm{region}}\sum_{kl}w_{ij,kl}\phi^0_{kl} = \Phi_\mathrm{region},\quad\mathrm{and}
	\end{equation}
	\begin{equation}
		\sum_{ij\ \mathrm{region}} \phi^{c'}_{ij} =  \sum_{ij\ \mathrm{region}}\sum_{kl}w_{ij,kl}\phi^{0'}_{kl} = \Phi'_\mathrm{region}
	\end{equation}
	by Equation \ref{eq:PSF_fluxconservation}. So, flux is conserved and we didn't have to make any assumptions about the data set used to make the calibration. With the linear model, this result appears trivial and perhaps not generalizable to the case where the measurements are nonlinear in the $\phi^0_{ij}$. Section \ref{sec:nonlinearity} explores what happens in that case. First, we demonstrate that the issue remains, and that the solution works, with some solar-like example cases.

	\subsubsection{Linear measurement model: summary and additional examples}\label{sec:linearmeasurement_solarexamples}

	We have demonstrated that when the resolution of the ground truth does not match that of the measurement, there can be an {\em apparent} miscalibration effect which scales with the resolution mismatch. The issue arises even when the measurement process is completely linear and there is no miscalibration in the instrument (i.e., $c_m=1$, $\phi^m_0=0$ in Eq. \ref{eq:linearmeasurement_model}). In that case a pixel-to-pixel comparison will show a non-unity slope implying a need for recalibration (Eq. \ref{eq:naivecal_noactualpsf}), even though the large-scale fluxes of the ground truth and measurement match (Eq., \ref{eq:linear_fluxconserving_curve}) and therefore there is no need for recalibration.
	
	This spurious apparent miscalibration effect is maximized when the ground truth has an uncorrelated (`salt and pepper') distribution between each pixel and its neighbors, and that has been used as the example case so far. If instead the ground truth is completely correlated over the size of the PSF (i.e., each pixel in the ground truth is completely correlated with its neighbors, out to the size of PSF), then it is trivial to show that the PSF does nothing and the effects of resolution mismatch (e.g., between measurement and ground truth) vanish: all points in the scatterplot will fall on the $y=x$ line. Although there is a major `salt and pepper' component to the solar flux distribution, it not the only component (see below). As a result, the degree of miscalibration due to resolution mismatch in the solar case will tend to fall between the maximal, all salt and pepper case, and the fully correlated case. 

	In this section, we consider two solar example cases: One using HMI data and the other using the MURaM simulation (non-sunspot regions only). The pixel scale is set to that of GONG, but for continuity with previous examples (and to make the effects are clearer), we begin with a 1 pixel wide PSF (GONG's average PSF, with seeing, is $\sim 3$ pixels wide). To make it clear that the issues are due exclusively to the PSF, the PSF is the {\em only} difference between `measurement' and `ground truth' in these examples (the terms are used in quotes for that reason), and it is applied with a linear convolution.
	
\begin{figure}
		\begin{center}\includegraphics[width=0.37\textwidth]{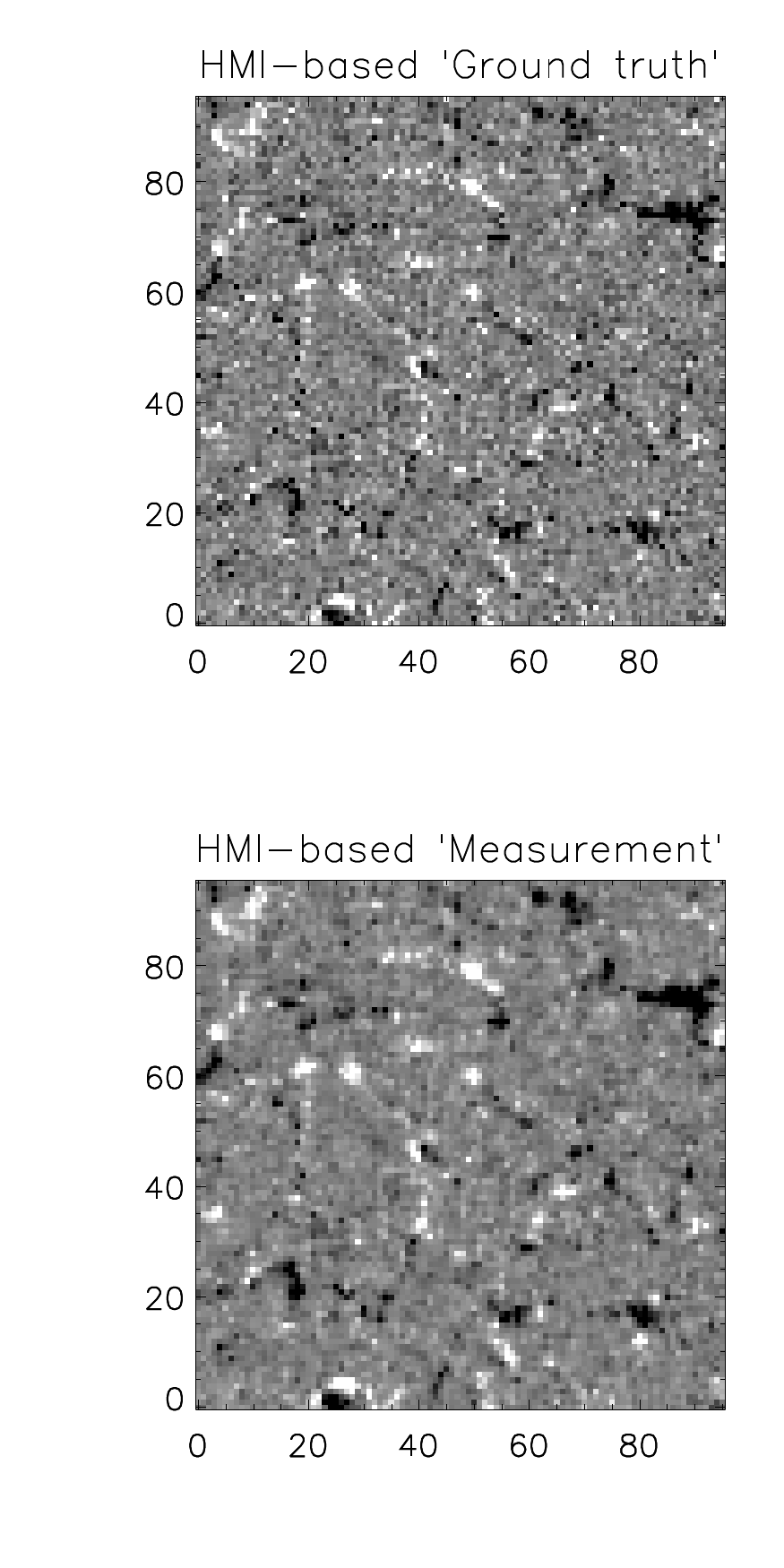}\includegraphics[width=0.63\textwidth]{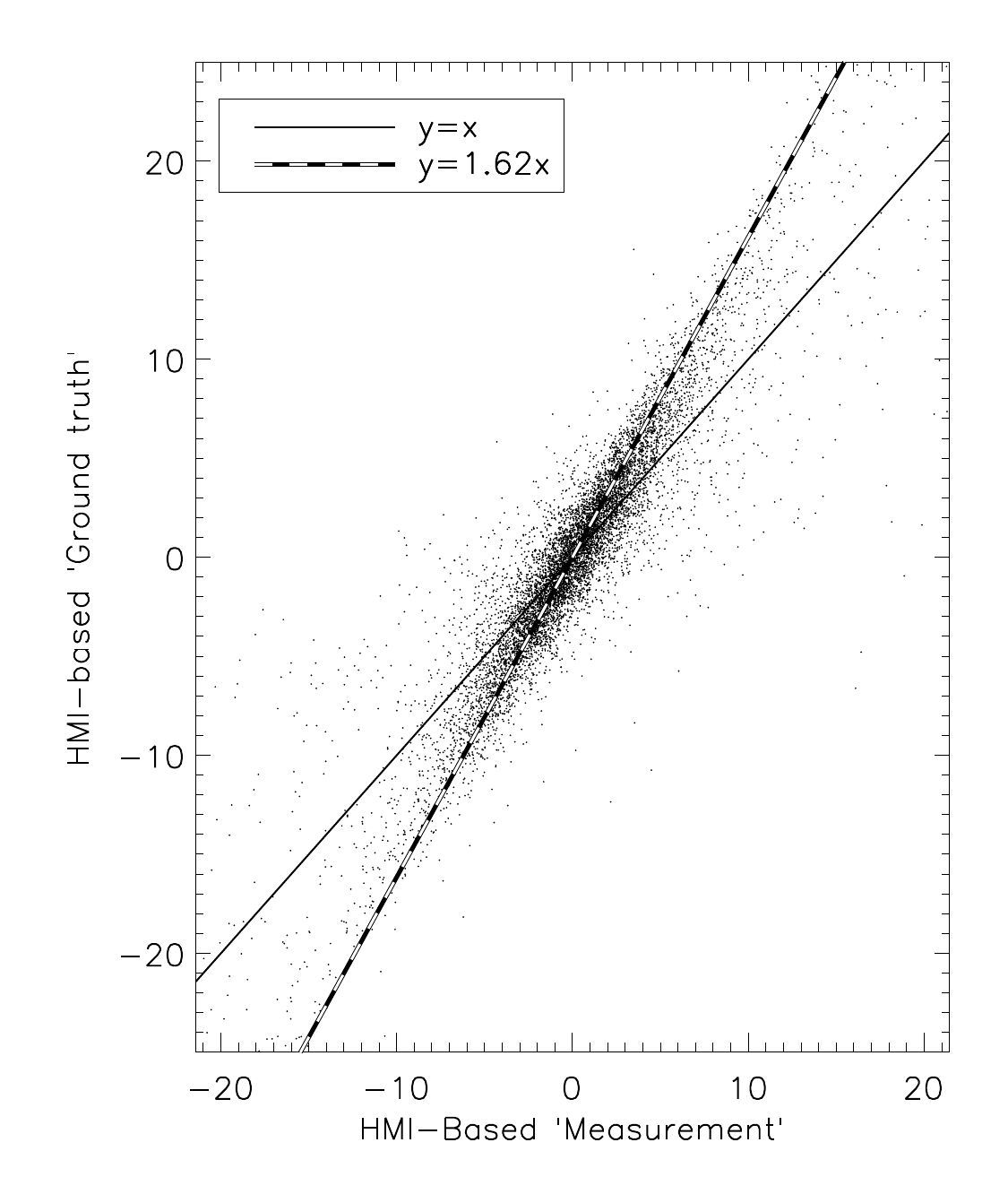}\end{center}
		\caption{Scatterplot (right) of `ground truth' (top left) vs. `measured' (bottom left) per-pixel fluxes for example case with solar-like level of GONG pixel-to-pixel correlation: an HMI magnetogram, resampled to the GONG pixel scale, is used as the `ground truth'. `Measurement' differs from ground truth in this example only in application of a linear, normalized PSF. As before, the per-pixel flux scatter plot is affected by the spatial resolution difference between measurement and ground truth, even though the net fluxes are not: The scatter plot for this solar example case appears almost identical to the uncorrelated case shown in Figure \ref{fig:simpleexample_scatterplot}.}\label{fig:HMIexample_scatterplot}
	\end{figure}

	Unlike the previous test case, the Sun is not completely uncorrelated. However, at GONG's pixel scale it is close enough to it, as far as this effect is concerned. This is illustrated in Figure \ref{fig:HMIexample_scatterplot}, which shows the point cloud resulting from using HMI data, reduced to GONG resolution, instead of the completely uncorrelated Gaussian random values. The slope is less than in the uncorrelated case ($\sim 1.4$ vs. $\sim 1.6$), but otherwise the point clouds are similar (We should perhaps point out that although HMI does also have a PSF, it is considerably smaller than the GONG pixels, so that at the GONG pixel scale it is essentially PSF free). This demonstrates that the `salt \& pepper' is the major component of the solar flux distribution at HMI scales. As a result, a similarly erroneous calibration will be obtained: As mentioned in Appendix \ref{app:PFSS}, no calibration factor is needed to make these two HMI-based magnetograms (one with the GONG PSF, and one without, both at GONG pixel scale) give the same extrapolations and open flux. The mismatched-resolution point cloud would have us inflate the magnetogram with PSF, which would instead bring those two extrapolations {\em out} of agreement.

\begin{figure}
		\begin{center}\includegraphics[width=0.37\textwidth]{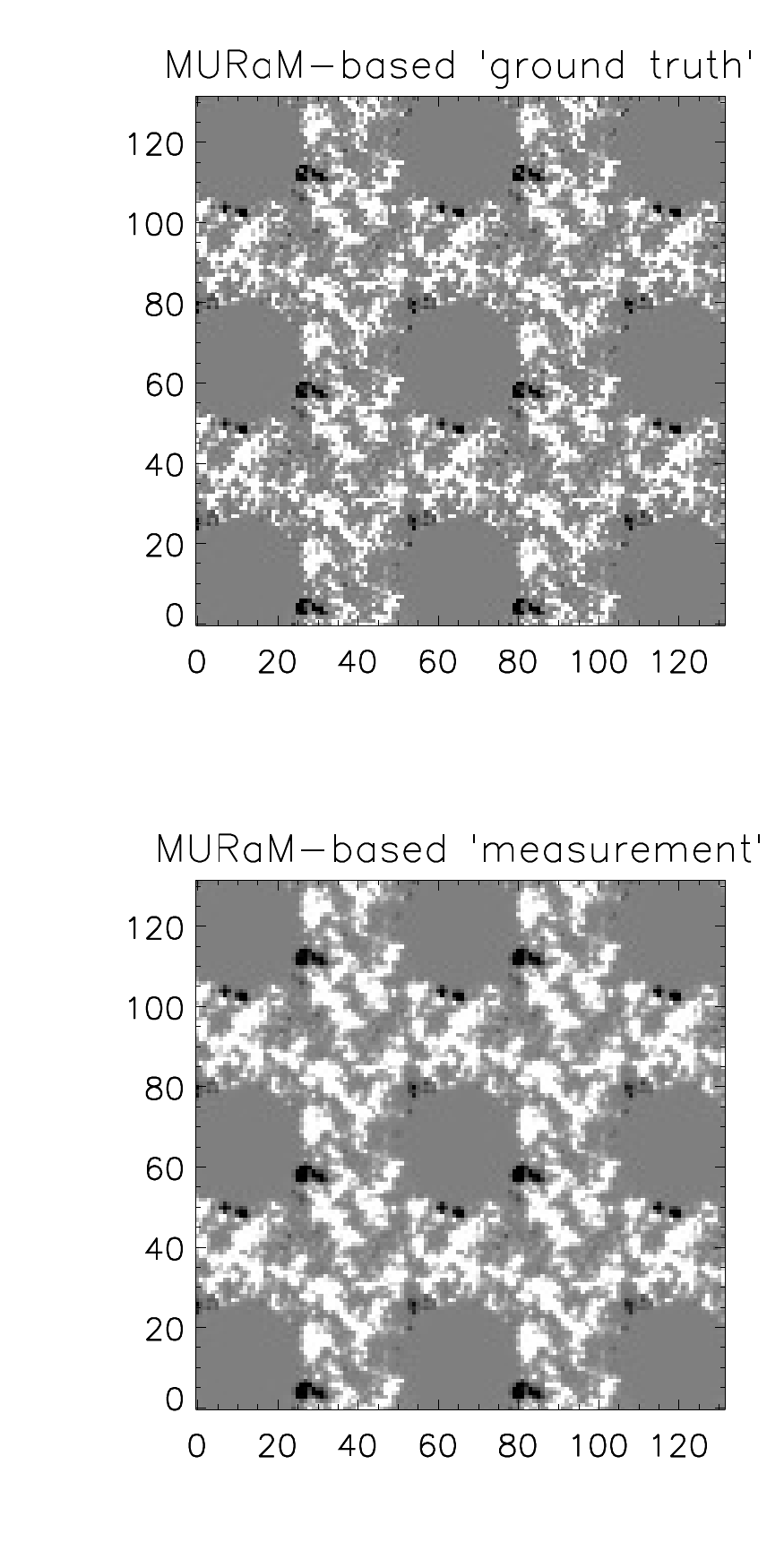}\includegraphics[width=0.63\textwidth]{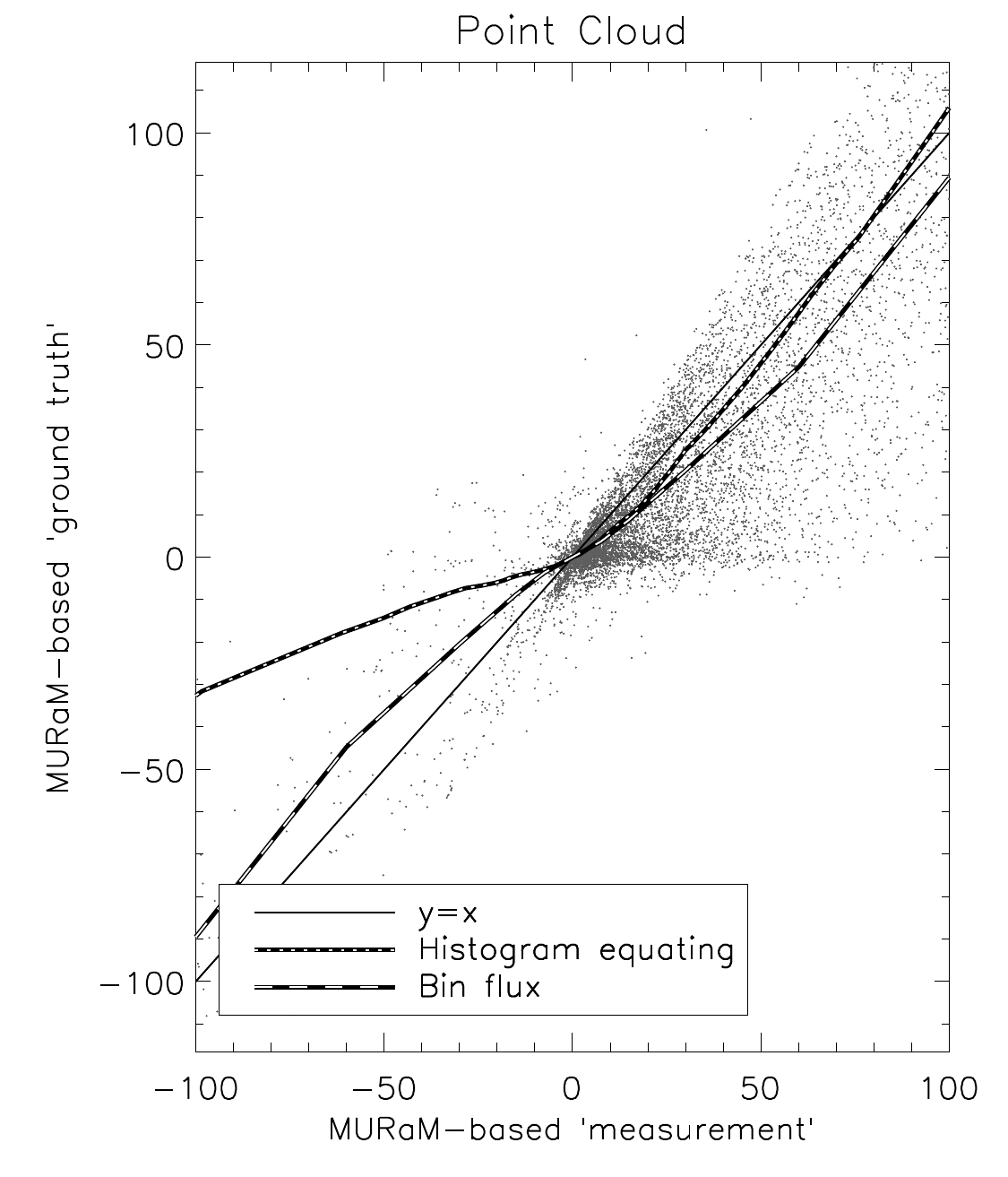}\end{center}
		\caption{Test case showing effects of linear 1-pixel wide PSF convolution on point clouds. In this example, the `ground truth' (top left) is from the MURaM simulation resampled to GONG pixel scale. The `measurement' (bottom left) is related to the ground truth only by application of a linear PSF; {\em all} of the differences shown by the scatter plot (right) are due to the PSF. The plot also shows fit curves based on histogram equating and on the curve fitting method (`Binned flux') described in Appendix \ref{app:fittingmethods}. The point cloud for this case is more complex than for the purely uncorrelated or HMI example cases; see discussion in text.}\label{fig:example_pointcloud_MURaM}
	\end{figure}

	We show the MURaM case in Figure \ref{fig:example_pointcloud_MURaM}. Because the spurious correlation effects arise in regions that are spatially uncorrelated (i.e,. random), we have masked out the sunspot from the pixel-to-pixel comparison. We have also tiled the simulation (which is quite small at GONG's pixel size) to flesh out the scatter plot. The results are more complex than in the pure random and HMI cases. This is primarily due to the flux imbalance in the simulation: it has a positive mean flux and a dramatic drop-off in the number of pixels (as a function of flux) for negative fluxes. As a result, the number of points in the scatter plot drop off abruptly for $y$-axis (`ground truth') fluxes less than zero. Notwithstanding this, the point cloud is consistent with a slope of $\sim 1.3$ rather than the $\sim 1.6$ predicted by the purely uncorrelated case: The positive part of the point cloud is well fit by Equation \ref{eq:naivecal_noactualpsf} with $K_{00}\approx 1/1.3$ rather than $1/1.6$ and offset ($\phi^0_0$) equal to the mean of the ground truth (this line is shown in Figure \ref{fig:example_pointcloud_MURaM}). This also suggests that the MURaM non-sunspot fields are somewhat less uncorrelated than in the real quiet sun. 
	
	The few points in the cloud with less than $\sim -10$ Gauss are associated with a handful of strong, spatially correlated negative regions near the sunspot, and they are likewise well fit by a linear relationship with the same slope but {\em negative} offset ($\phi^0_0$). When these positive and negative regions are part of the same point cloud, a curve fit to them will have a `kink' at the origin: the resulting weak-flux slope of the curve will be small. In reality, this is not due to differing slopes in the cloud (the calibration relationship), but to combining two clouds (i.e., two flux populations: one for the small negative-dominant regions, one for the rest) with different offsets. If the spatial correlation of the negative dominant points in the cloud is removed, by randomly shuffling the pixel locations in the ground truth, the kink goes away and the point cloud is well fit by Equation \ref{eq:naivecal_noactualpsf} with $K_{00}\approx 1/1.6$ and $\phi^0_0$ equal to the mean of the ground truth (not shown).

\begin{figure}
		\begin{center}\includegraphics[width=0.37\textwidth]{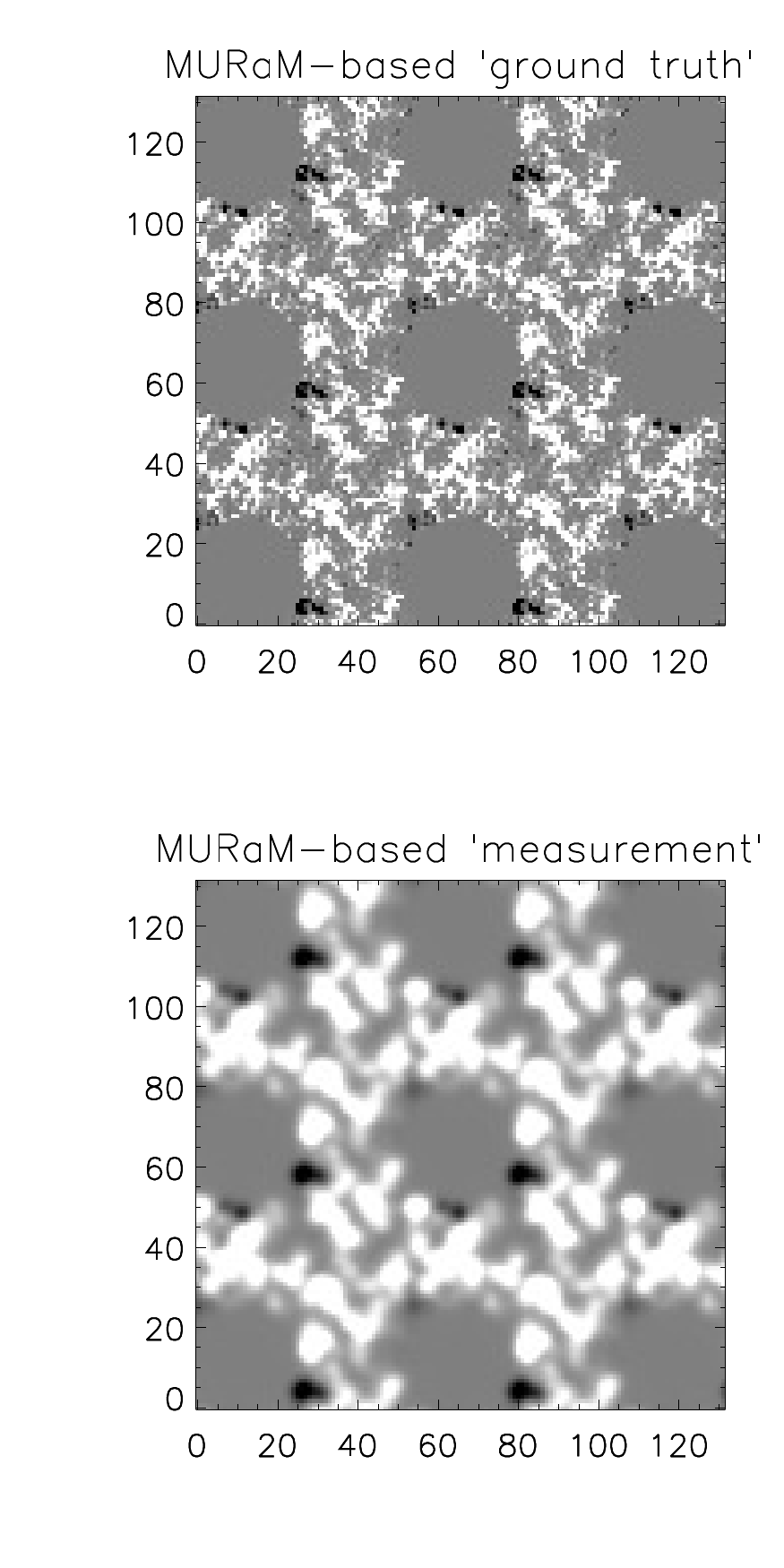}\includegraphics[width=0.63\textwidth]{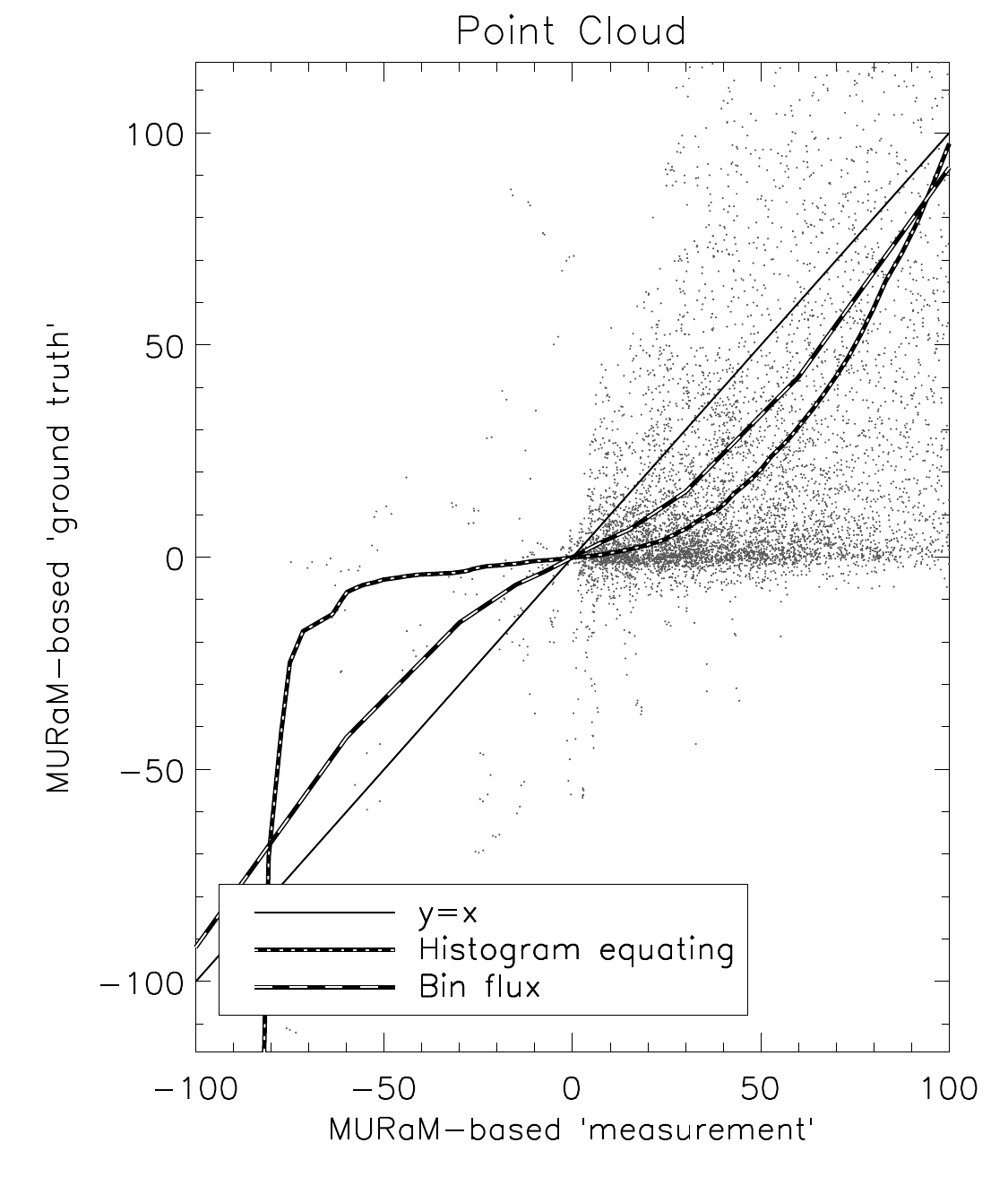}\end{center}
		\caption{Test case showing effects of linear 3 pixel wide PSF convolution on point clouds. In this example, the `ground truth' (top left) is from the MURaM simulation resampled to GONG pixel scale. The `measurement' (bottom left) is related to the ground truth only by application of a linear PSF; {\em all} of the differences shown by the scatter plot (right) are due to the PSF. The plot also shows fit curves based on histogram equating and on the curve fitting method (`Binned flux') described in Appendix \ref{app:fittingmethods}. With the PSF resolution difference in play, there is no clear relationship between the `measurement' and the `ground truth', illustrating the need for resolution matching.}\label{fig:MURaM_example_pointcloud_3px}
	\end{figure}

	The large scatter and lack of consistent correlation shown by these point clouds, caused by just a 1 one-pixel wide PSF, makes them difficult to justify as a calibration relationship. These issues are caused by the PSF alone, since the PSF is the only measurement effect present in these example cases. This is made worse when a 3-pixel wide PSF (GONG's PSF is roughly this size) is used instead of a 1-pixel wide one. For the MURaM example case, the `kink' and the amount of scatter in the point clouds is very large: see Figure \ref{fig:MURaM_example_pointcloud_3px}. No clear relationship between measurement and ground truth, upon which a calibration could be based, is evident; it has been effaced by the PSF.

	Figures \ref{fig:MURaM_example_pointcloud_3px} and \ref{fig:example_pointcloud_MURaM} also show that an S-shape histogram equating curve is not necessarily a sign of `saturation' of the magnetographs, as is sometimes claimed. This synthetic example is entirely linear in the fluxes, and the S-shaped curve is entirely a result of the different resolutions.

\begin{figure}
		\begin{center}\includegraphics[width=0.37\textwidth]{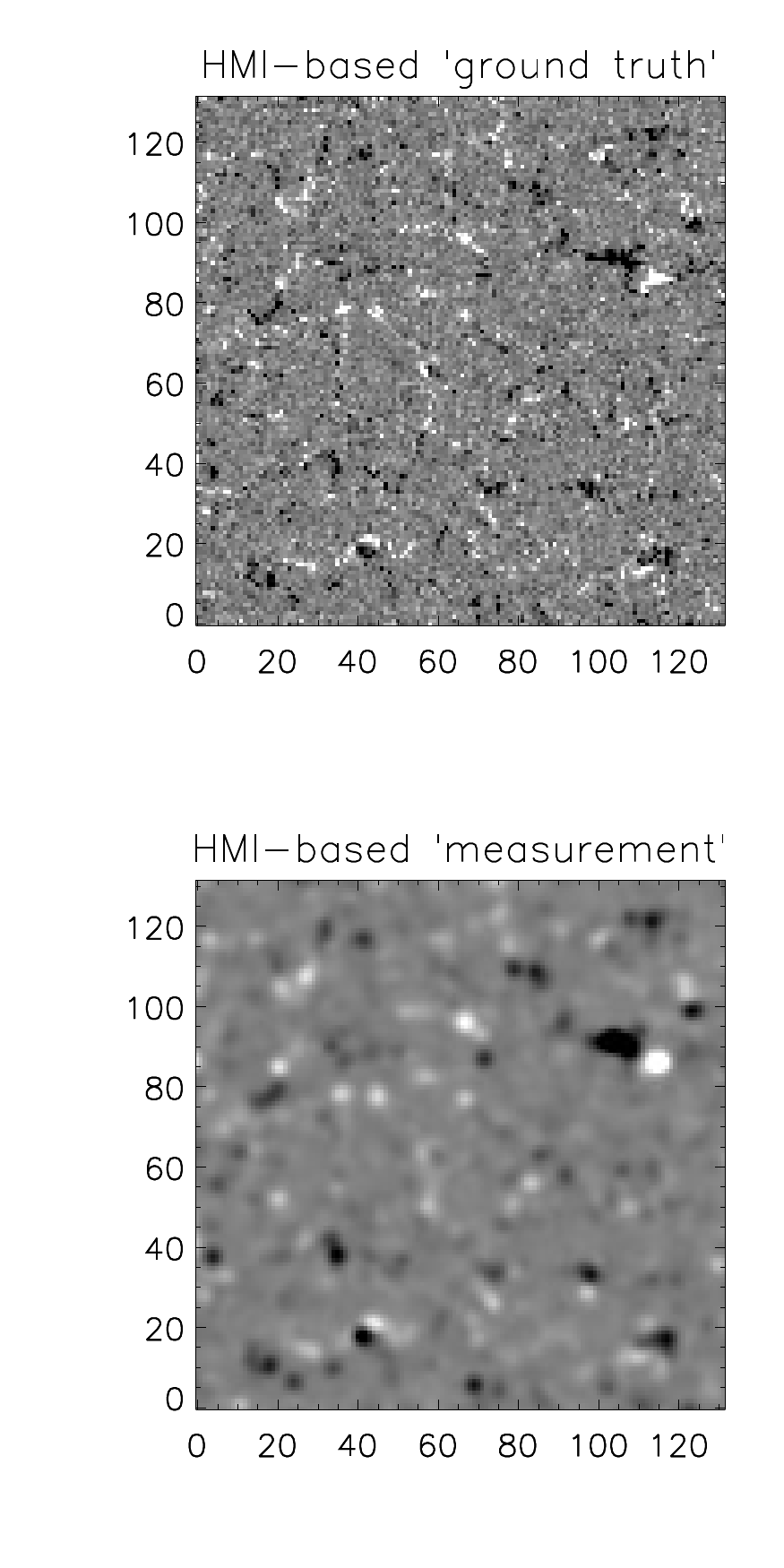}\includegraphics[width=0.63\textwidth]{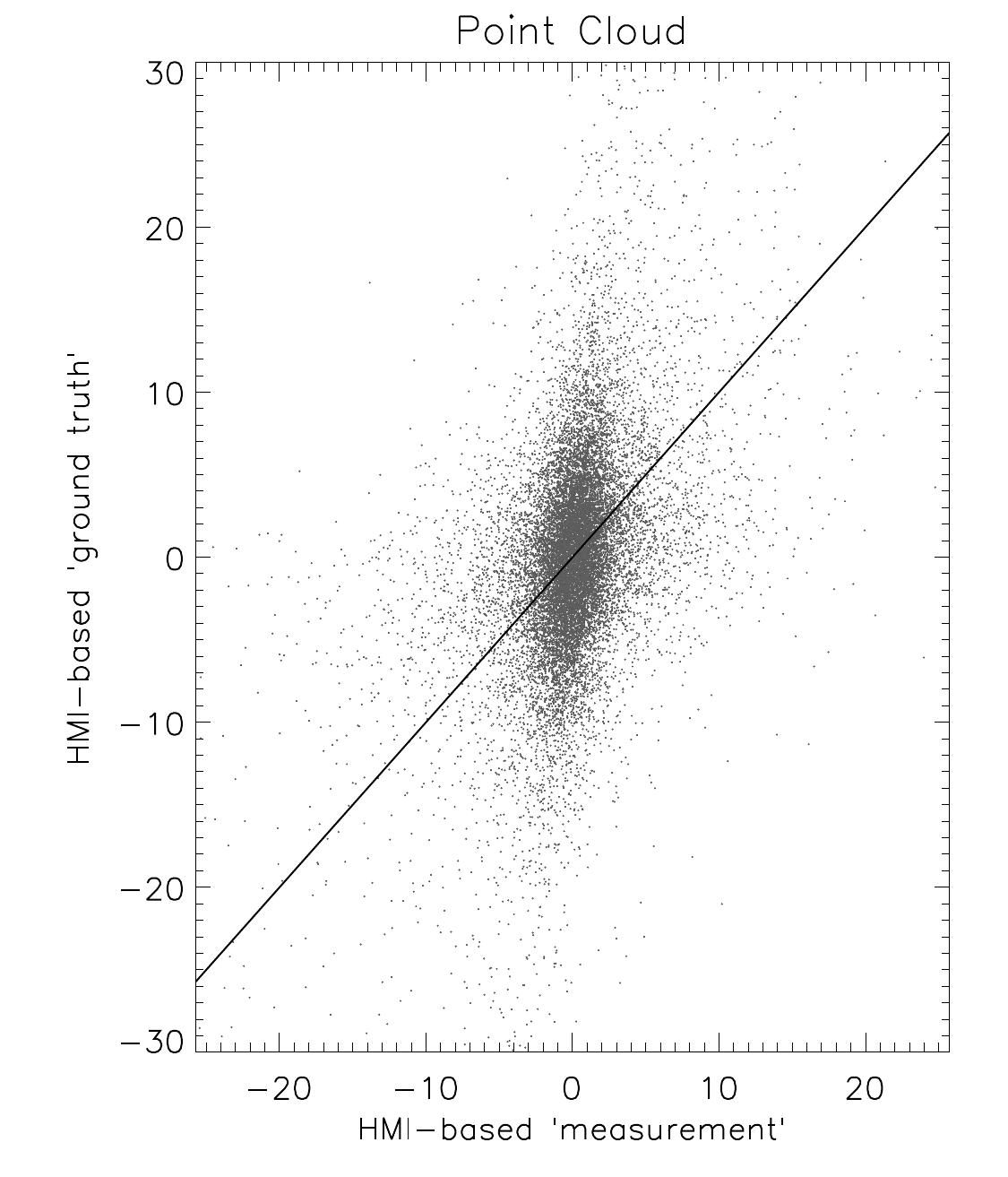}\end{center}
		\caption{Test case showing effects of linear 3 pixel wide PSF convolution on point clouds. In this example, the `ground truth' (top left) is from an HMI magnetogram binned down to GONG pixel scale, and the `measurement' (bottom left) differs from the ground truth only in the application of the PSF. The scatter (plotted on right) is large and slope(s) poorly defined, so no curve fits or slopes (other than $y=x$) are shown, to avoid cluttering the figure and distracting the eye.}\label{fig:HMI_example_pointcloud_3px}
	\end{figure}

	This is not restricted to the MURaM case, and issues would persist even for a ground truth without its peculiarities: for the HMI example with 3 pixel PSF (Figure \ref{fig:HMI_example_pointcloud_3px}), there is no kink, but the scatter in the point clouds is still very large and the slope is poorly defined: different fitting methods give very different curves. The cloud does appear to show the $1/K_{00}$ slope (for the 3 pixel wide PSF, this is a factor of $\sim 14$) predicted for purely uncorrelated ground truth.

\begin{figure}
		\begin{center}\includegraphics[width=0.37\textwidth]{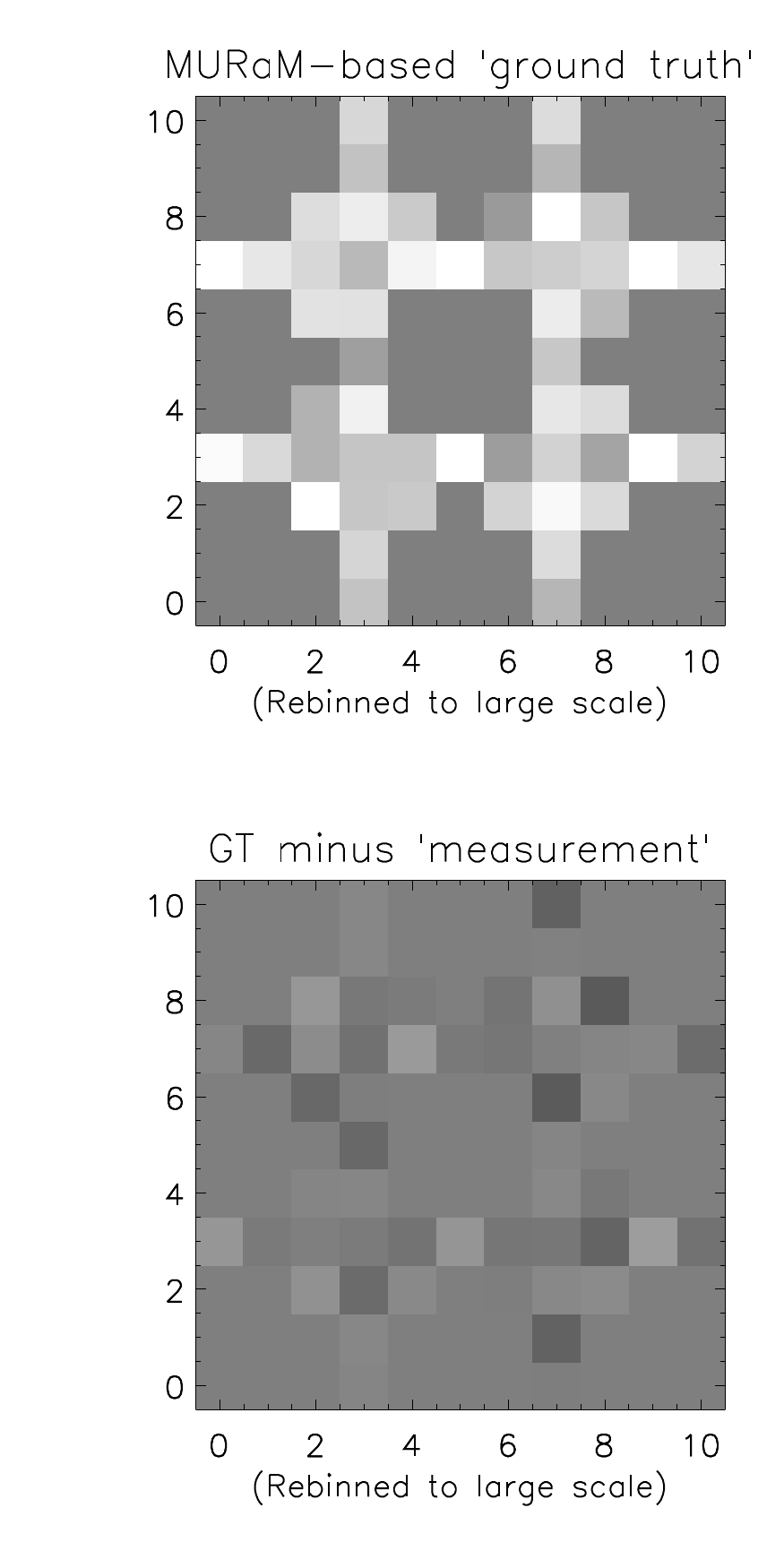}\includegraphics[width=0.63\textwidth]{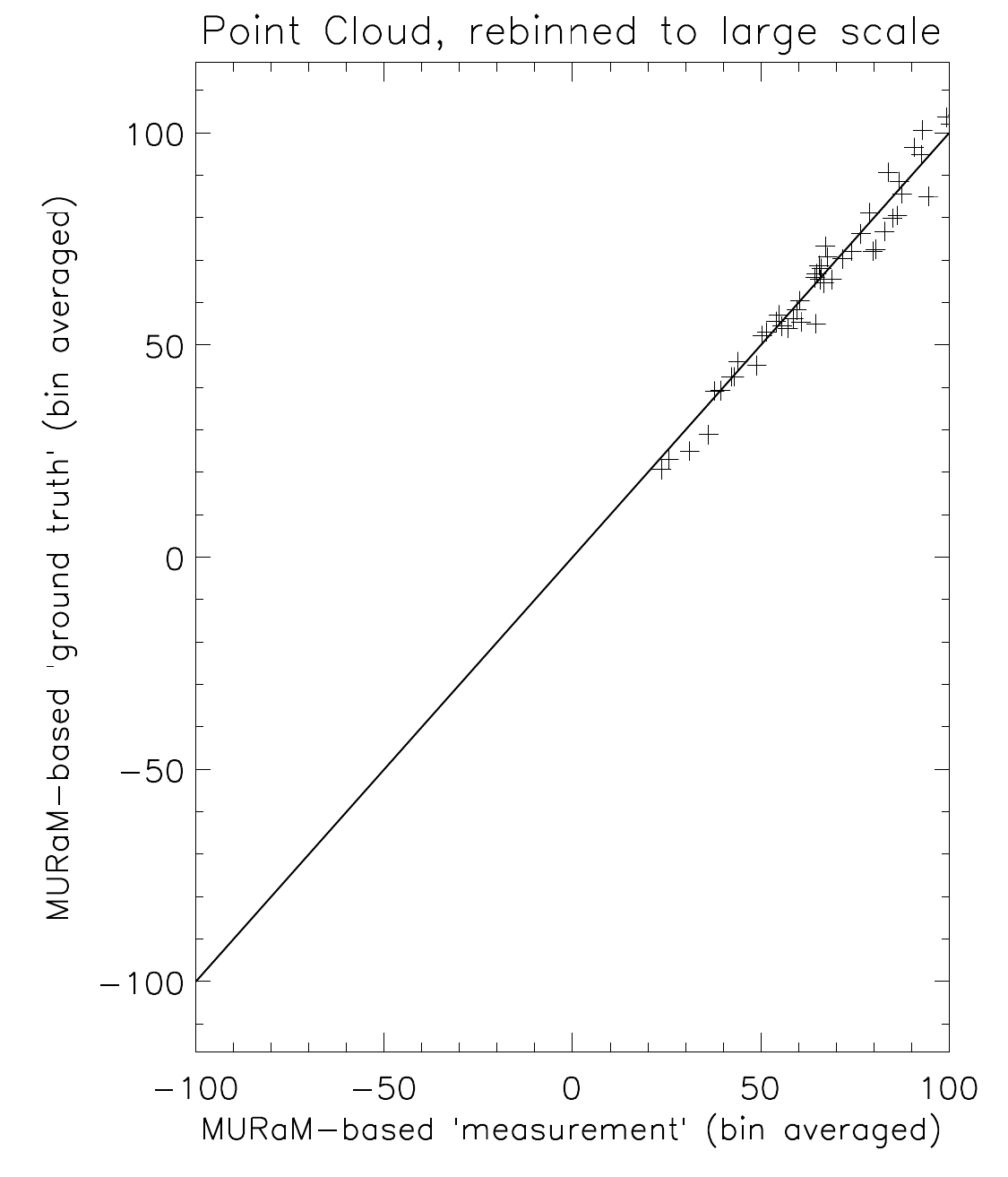}\end{center}
		\caption{Test case showing effects of linear 3 pixel wide PSF convolution on large-scale fluxes. In this example, the `ground truth' (top left) is from the MURaM simulation resampled to GONG pixel scale. The `measurement' is related to the ground truth only by application of a linear PSF; {\em all} of the differences shown by this point cloud are due to the PSF. The bottom left panel shows the difference between ground truth and measurement. Otherwise, the only difference between this figure and Figure \ref{fig:MURaM_example_pointcloud_3px} is that the measurement and ground truth have been rebinned (12 by 12) before plotting and forming the point cloud (right panel). The dramatic differences shown in that figure's scatter plots completely vanish when the measurement and ground truth are rebinned to large scale.}\label{fig:MURaM_example_pointcloud_3px_largescale}
	\end{figure}

	All of these issues are caused by the resolution mismatch between the measurement and the ground truth and the lack of complete correlation over scales the size of the PSF: The PSF is the only difference between `measurement' and `ground truth' in this example and (as previously noted) the PSF does {\em nothing} if there is complete correlation over regions the size of the PSF. The dramatic differences shown in these point clouds all vanish (including the `kink') when the measurement and ground truth are rebinned to large scales and then compared (see Figure \ref{fig:MURaM_example_pointcloud_3px_largescale}): the slope of the large-scale scatterplot is unity with minimal scatter. Therefore, no recalibration in this example is needed, just as Equation \ref{eq:linear_fluxconserving_curve} would predict. However, as before, the effects of the mismatch on a calibration curve do {\em not} vanish when `calibrated' measurement and ground truth are compared at larger scales. 

\begin{figure}
		\begin{center}\includegraphics[width=0.37\textwidth]{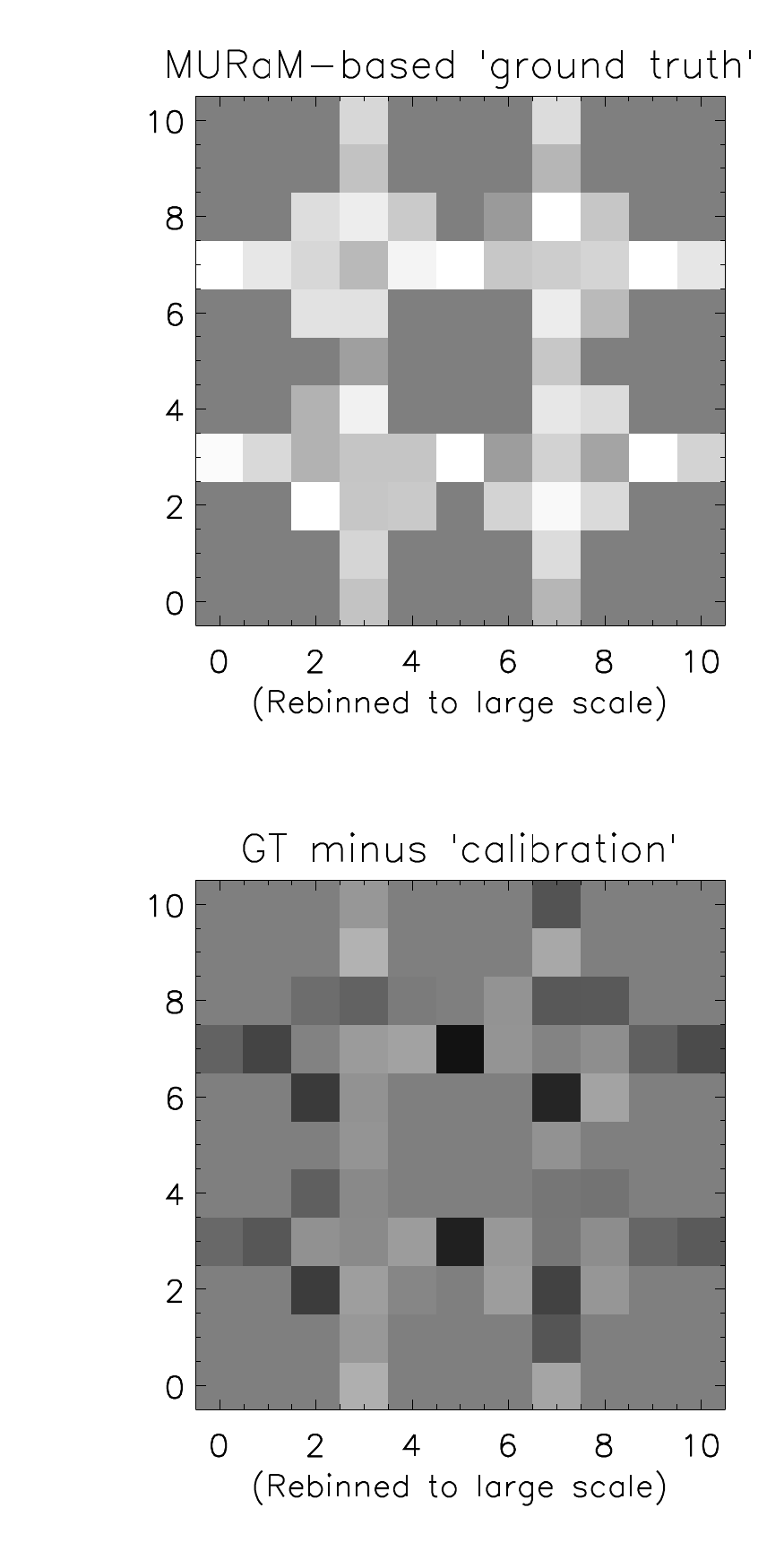}\includegraphics[width=0.63\textwidth]{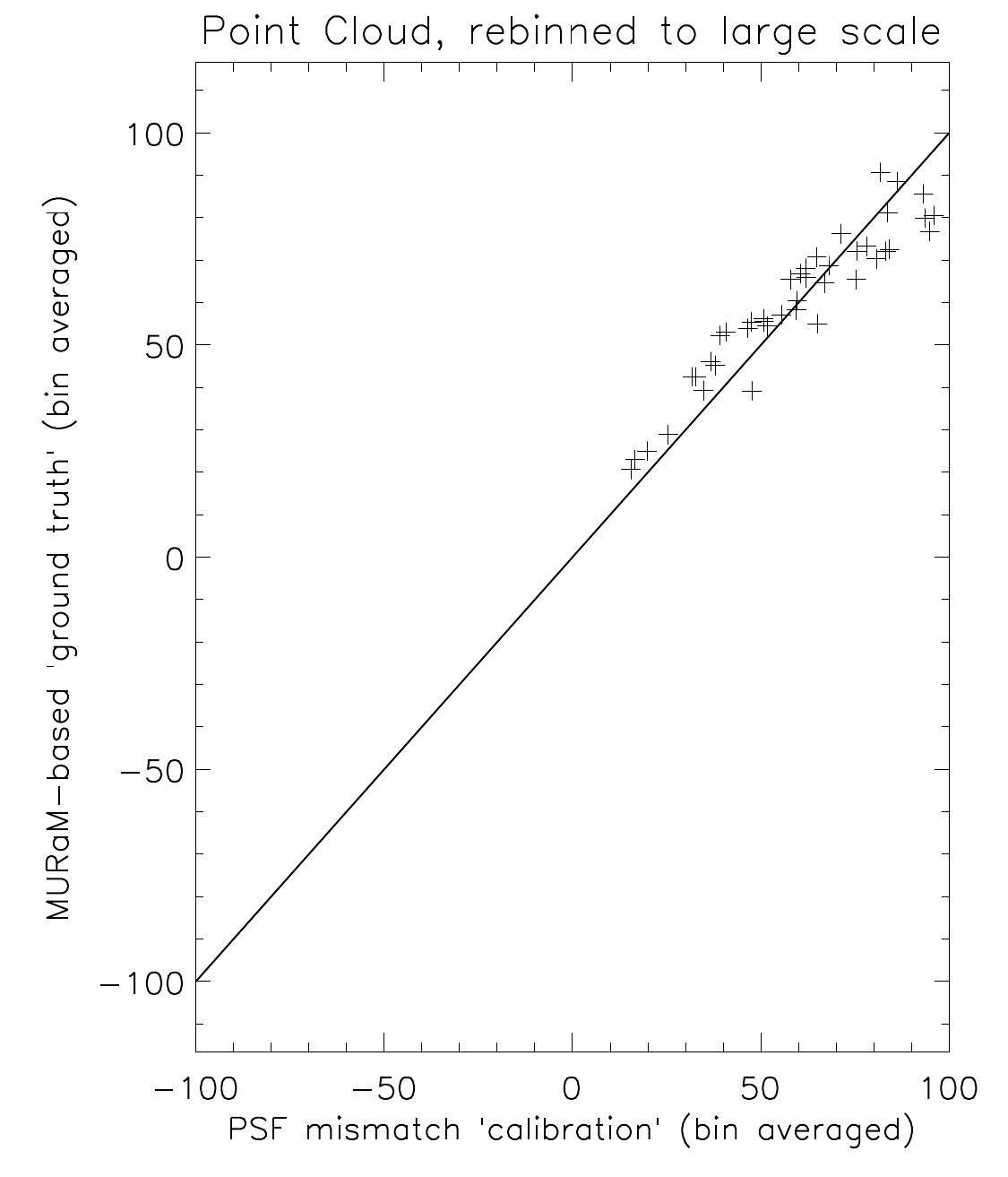}\end{center}
		\caption{Effects of applying the `calibration' curve fit shown in Figure \ref{fig:MURaM_example_pointcloud_3px} to the `measurement' shown in the same figure. This figure is identical to Figure \ref{fig:MURaM_example_pointcloud_3px_largescale} except in the application of the `calibration' curve to the measurements, and it can be seen that it makes the large-scale fluxes worse; compare lower left and right panels of the figures and see discussion in text.}\label{fig:MURaM_example_pointcloud_3px_largescale_miscal}
	\end{figure}

	Figure \ref{fig:MURaM_example_pointcloud_3px_largescale_miscal} demonstrates this by applying a `calibration' derived from Figure \ref{fig:MURaM_example_pointcloud_3px} back on the measurements used to make the calibration (as in Figure \ref{fig:groundtruthpsf_badcal_netflux}), rebinning to large scales, and then making a point cloud of the results. In this case, the `kink' and the steeper slope at high flux partly counteract each other, and the small size of the MURaM simulation (compared to the GONG pixels) means that the large-scale fluxes are all fairly similar: the simulation is tiled so that there are more than a few points in the cloud, but they are all sampling the same regions (with different large-scale pixel centers). As a result, the slope of the (large-scale) point cloud is not too different from one, although still worse than if no calibration had been applied at all. However, the scatter of the `calibrated' large-scale fluxes is still larger than those with no calibration; the large-scale fluxes are therefore made worse by the `calibration', albeit not in a systematic fashion (compare especially the middle panels of Figures \ref{fig:MURaM_example_pointcloud_3px_largescale} and \ref{fig:MURaM_example_pointcloud_3px_largescale_miscal}).

	This changes dramatically if that calibration is applied to a region with a different flux distribution than the MURaM simulation. If the region consists mostly of measured fluxes over $\sim 100$ Gauss, then the high slope in that part of the point cloud (Figure \ref{fig:MURaM_example_pointcloud_3px}) will lead to overestimation of the flux. If, on the other hand, the region consists of measured fluxes under $\sim 100$ Gauss, then the small slope of the calibration curve in that part of the point cloud will cause the `calibrated' data to underestimate the `true' fluxes (i.e., the fluxes used to make the `measurements' in this PSF-only test case). This is the case across much of the real sun, and would still be the case even if GONG (or HMI) are underestimating their fluxes by a factor of three. 

	It is also worth pointing out that the curve has not removed the resolution differences caused by the PSF: Although the axes of the per-pixel point cloud are equalized, their large scatter remains, and the `calibrated' images are still smoother than the ground truth (`sharpening' them would require a multidimensional convolution).

	We have demonstrated that a 1-dimensional calibration curve can't correct for the resolution differences caused by a PSF, and that attempting to do so results in incorrect `calibrated' large-scale fluxes. The large-scale flux issue is resolved by reducing the ground truth to the same resolution, including the instrument PSF and seeing, as the measurements. In the linear case, this is nearly trivial, as equations \ref{eq:resmatch_linearcomparison} through \ref{eq:resmatch_calcurve} demonstrate. In the interests of brevity, additional graphical demonstration is deferred until the nonlinear case is considered (in Section \ref{sec:nonlinearity}). 

	Lack of resolution matching is likely a significant factor in the range of results  found in magnetograph comparisons in the literature. Appendix \ref{sec:resmismatch_literature} gives a brief review. In short, some publications do not give any indication that they have matched resolutions. These publications often also use the `histogram equating' comparison method which does not require an explicit pixel-to-pixel correspondence. It might be thought that this lack of correspondence makes it immune to the resolution mismatch issue we have described, but, as we show in Appendix \ref{sec:histogram_equating}, it is affected in a very similar way. Some other publications \citep[e.g.,][]{2010LambEtal_ApJ720_140, 2012LiuEtal_SoPh279_295L, 2013PietarilaEtal_SoPh282_91} have done resolution matching in their comparisons, and found that it helps their comparisons, demonstrating that this means of correction already has a peer reviewed track record.

	\subsection{Pixel windowing, the PSF, and the effects of nonlinearity}\label{sec:nonlinearity}

	We have considered the `ideal' case where the effect of PSF on the magnetogram can be treated as a linear convolution: the observed fluxes are a linear function of the ground truth fluxes only. In reality, the fluxes inferred from observables (polarized spectra) produced by the radiative transfer process, which we denote by $\mathbf{O}(x,y,\lambda)$. The radiative transfer process depends on the ground truth fluxes directly via the Zeeman effect, but also on the plasma temperature, velocity, and density. The radiative transfer process is not linear in {\em any} of these parameters, and the nonlinearity in the observables' dependence on the flux is reinforced by the correlation between the plasma temperature, density, velocity, and flux.
	
	The observing process integrates these observables against wavelength response functions \citep[we have listed these functions for GONG in the first paper of this series,][]{PlowmanEtal_2019I}, point spread functions, $p(x-x',y-y')$ (where $x'$ and $y'$ are coordinates on the instrument focal plane), and pixel `window' functions $\theta(x'-x_i,y'-y_j)$, which are in the form of square or rectangular top hat functions: 1 if $|x'-x_i| < \delta_x/2$ and $|y'-y_j| < \delta_y/2$ and zero otherwise ($\delta_x$ and $\delta_y$ are the pixel sizes in $x$ and $y$, while $x_i$ and $y_j$ are the pixel center locations). The observable recorded by a given pixel, $\mathbf{O}_{ij}$, can then be written:

	\begin{equation}
		\mathbf{O}_{ij} = \int_{-\infty}^{\infty}\int_{-\infty}^{\infty}\theta(x'-x_i,y'-y_j)\int_{-\infty}^{\infty}\int_{-\infty}^{\infty}p(x-x',y-y')\mathbf{O}(x,y)dxdydx'dy'
	\end{equation}
	This can be rewritten in terms of an overall pixel `spatial response function', $R_{ij}$, which encapsulates how the pixel responds to light incident on the instrument (in place of the pixel-wise PSF convolution in earlier sections), as a function of sky angle:
	\begin{equation}\label{eq:spatialresponse}
		R_{ij}(x,y) \equiv  \int_{-\infty}^{\infty}\int_{-\infty}^{\infty}\theta(x'-x_i,y'-y_j)p(x-x',y-y')dx'dy',
	\end{equation}
	This can be recognized as the convolution of the pixel window function and the PSF. In terms of this, the pixel observables are the plane-of-sky observables integrated against each pixel's spatial response function.
	\begin{equation}\label{eq:pixelresponse}
		\mathbf{O}_{ij} = \int_{-\infty}^{\infty}\int_{-\infty}^{\infty}R_{ij}(x,y)\mathbf{O}(x,y)dxdy
	\end{equation}

	These pixel observables are the recorded by the detector and `inverted' to produce the flux measurements, as represented by the functions $o^{-1}_{ij}$ \citep[For GONG, we have described this inversion in][]{PlowmanEtal_2019I}:

	\begin{equation}
		\phi^m_{ij} = o^{-1}_{ij}(\mathbf{O}_{ij}).
	\end{equation}
	We use a lower-case $o^{-1}$ as a reminder that these are not a true inverse due to loss of information in the forward problem. Now we want to relate them back to the ground truth values for each pixel, $\phi^\mathrm{GT}_{ij}$. As previously noted, the measurements depend not on a single value but on those at many points in the simulation volume. How to distill them all down to just one flux for each pixel?

	Consider the case where there is no PSF (or, more properly, the PSF is a Dirac delta function). There, the answer is quite clear: $\phi^\mathrm{GT}_{ij}$ should be the ground truth fluxes (or fields) integrated over the pixel window function.
	
\begin{figure}
		\begin{center}
			\includegraphics[width=0.3\textwidth]{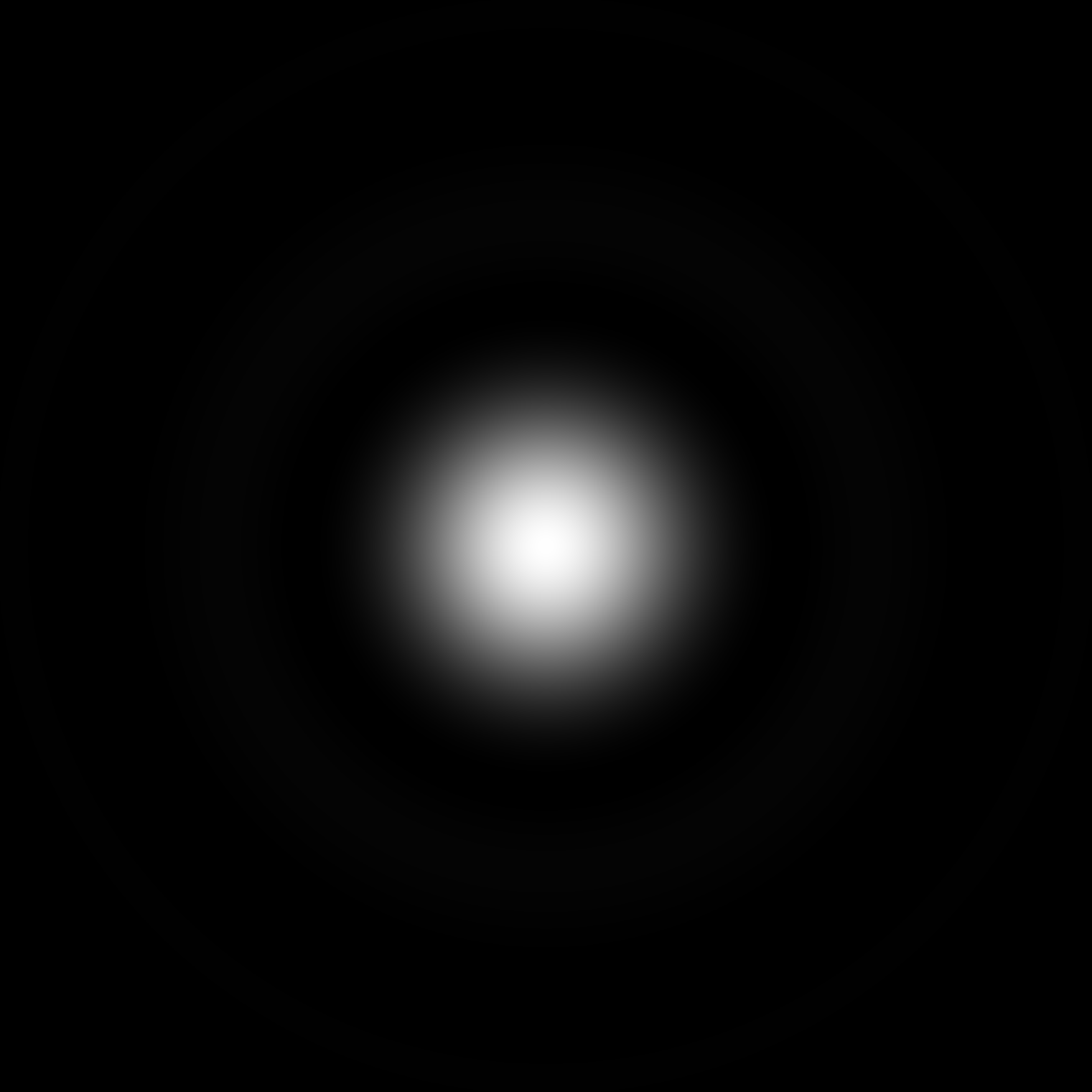}
			\includegraphics[width=0.3\textwidth]{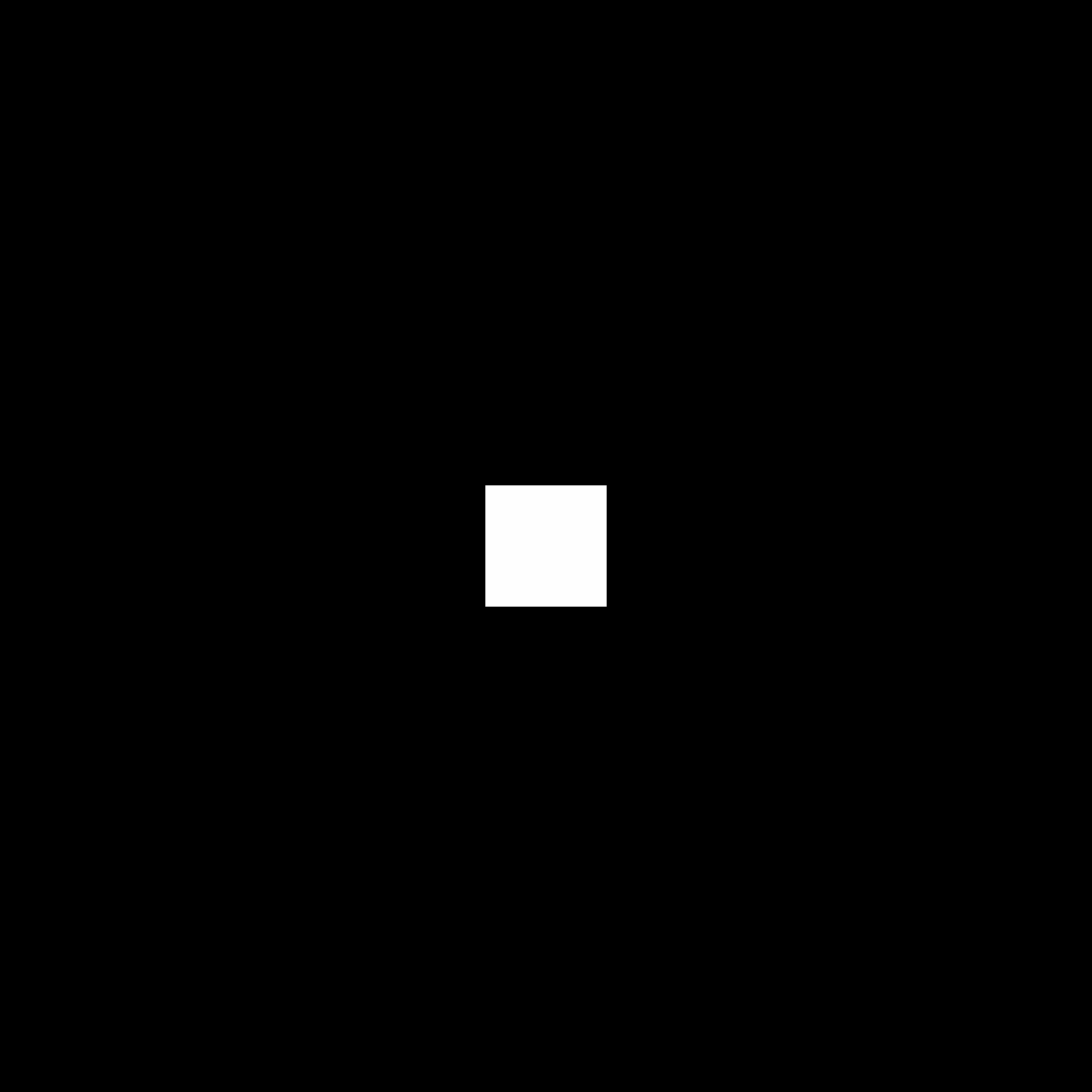}
			\includegraphics[width=0.3\textwidth]{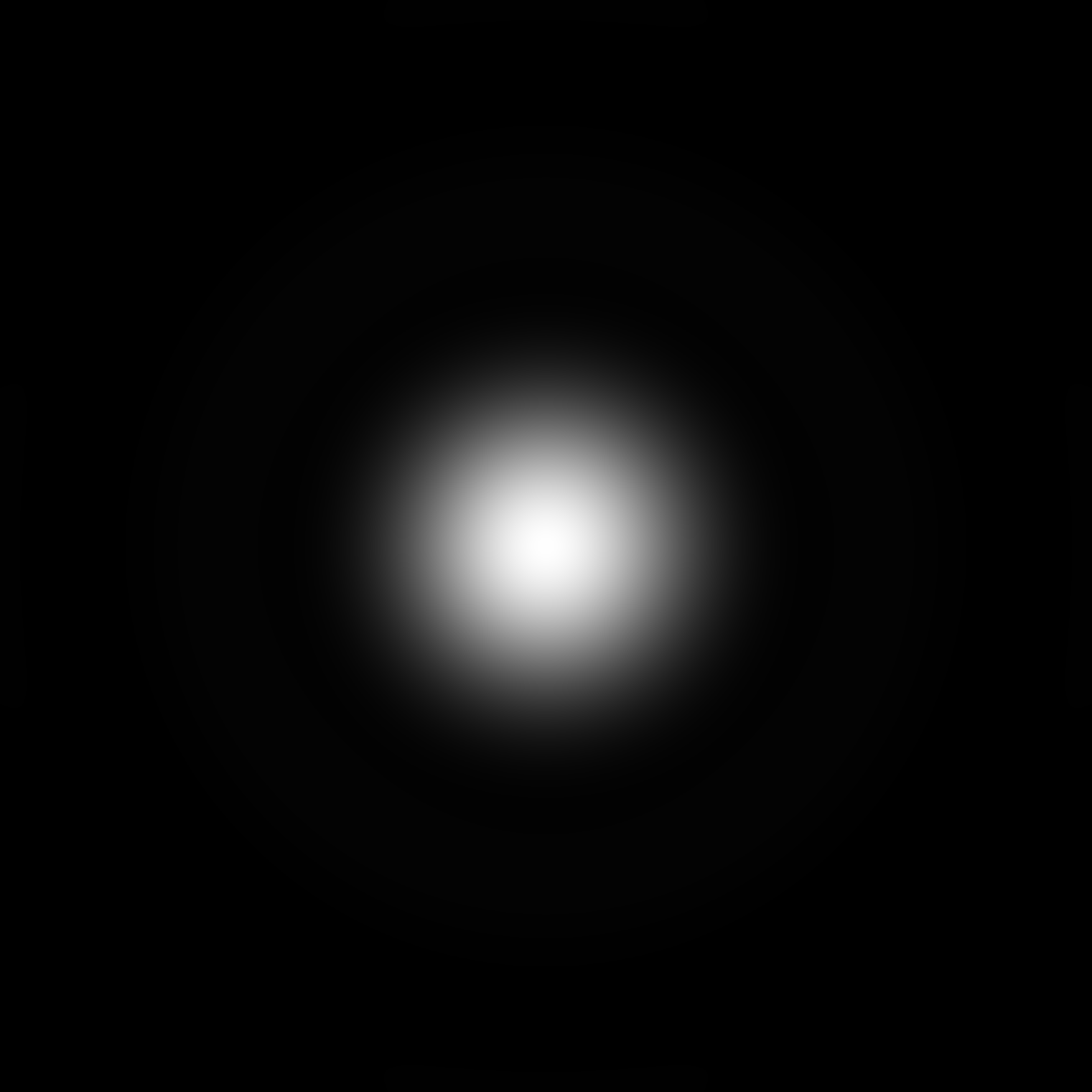}
		\end{center}
		\caption{The overall spatial response function obtained by combining the instrument PSF and the pixel window. Left: the PSF; center: The pixel window function; Right: the overall spatial response function, which is the convolution of the two.}\label{fig:pixelsampling}
	\end{figure}

	In general, however, the mathematics of the transformation (Equation \ref{eq:pixelresponse}) make no distinction between the PSF and the pixel window function -- the dependence is on the pixel spatial response function $R_{ij}$, not the PSF and window functions separately. There are no hard edges left when the PSF and the pixel window functions are combined. The instrument simply samples the observables ($O(x,y)$) at regular intervals, with highly overlapping sampling functions given by the $R_{ij}$ (Equation \ref{eq:spatialresponse}). The only remaining dependence on the pixel grid is in the spacing between the spatial sampling points, and each of these spatial sampling functions have large overlaps: there are no `dividing lines' between a pixel and its neighbors, as far as their response to the solar observables are concerned. This is illustrated in Figure \ref{fig:pixelsampling}.
	
	Similarly, although images are often displayed with hard-edged pixels on computer screens, this is only a convenient fiction employed for visualization purposes. It does not represent how the instrument is sampling its source on the sky.

	The goal of calibration should be to more faithfully intrepret what the measurements are telling us about the sun, not to force them to tell us something else. These measurements are very specifically telling us about the fluxes averaged over the entire spatial response function, not just the central pixel: the outlying pixels contribute to the measurement in exactly the same fashion, mathematically, as the central one. While the outlying pixels have less weight {\em individually} than the central one, combined they have more: in GONG's case, roughly 90\% of the weight of the spatial response function comes from outside the central pixel.

	The overall takeaway is that if it is appropriate to use
	\begin{equation}
		\phi^\mathrm{GT}_{ij} = \int \theta_{ij}(x',y')B_z(x',y')dx'dy'
	\end{equation}
	in the case where the pixel response is the same as the window function (i.e., when the PSF is a delta function), then it is also appropriate to use
	\begin{equation}
		\phi^\mathrm{GT}_{ij} = \int R_{ij}(x',y')B_z(x',y')dx'dy'
	\end{equation}
	in the general case. On the other hand, if nonlinearity makes it inappropriate to use the pixel response function in the general case, then it is also inappropriate to use the pixel window function even in the case where the PSF is a delta function. Whether or not nonlinearity makes it inappropriate to compute the $\phi^\mathrm{GT}_{ij}$ from an average (weighted or not) at all is a separate question. The answer to this question depends on the following considerations:
	\begin{itemize}
		\item If, for any given measured value, the point cloud has a single well-defined ground truth value (i.e., the degree of scatter is minimal), then a curve fit to the relationship will restore the ground truth fluxes from the measurements: the forward mapping is well-defined, so the reverse mapping is as well.
		\item On the other hand, if there is not a single well-defined ground truth (e.g., the degree of scatter is significant) for each measured value, then it may not be possible to restore the ground truth from the measurement using the point cloud. Moreover, it is not possible to determine whether or not this is possible based on the appearance of the point cloud -- even if it shows a clear linear correlation. The linear measurement examples in the previous sections demonstrate this.
		\item If there is scatter in the point clouds, their flux conservation (or lack thereof) can be investigated directly by seeing whether or not the slope of the scatter plots change when they are rebinned to larger scales -- if the same slope is obtained regardless of the spatial scale of the data, then it is due to the intrinsic instrument calibration and not these scale effects.
	\end{itemize}
	
	\noindent We investigate this next, with a variety of examples.

	\subsubsection{Pixel windowing, the PSF, and nonlinearity: examples}\label{sec:psfnonlinearity_examples}

	We now consider how these spatial (i.e., plane-of-sky) resolution effects are changed when they enter the measurement process in a nonlinear way, specifically the GONG measurement principal \citep[described in][]{PlowmanEtal_2019I}. To isolate these spatial resolution effects from other parts of this process, we assume the wavelength shift in Stokes $I \pm V$ along each line of sight is proportional to the line of sight flux/field strength ($B_\mathrm{LOS}$) as we describe in \cite{PlowmanEtal_2019I} -- the procedure is as follows:
	\begin{itemize}
		\item Begin with a two-dimensional `ground truth' magnetogram as before, 
		\item convert it to a set of wavelength shifts ($0.03205$ \AA\ per kiloGauss),
		\item assume a reference spectrum and convert shifted wavelength profiles to intensities using the GONG wavelength response functions  from \cite{PlowmanEtal_2019I},
		\item downsample to GONG's resolution (including PSF),
		\item invert using the equations from \cite{PlowmanEtal_2019I}.
	\end{itemize}

	The resulting inverted fluxes are then compared with the `ground truth' (after also resampling it to the same resolution as the GONG output, including PSF). In the discussion of the linear measurement model (Section \ref{sec:spatialresolution}), the measurement and ground truth were assumed to have the same pixel grid for simplicity, but now the `native' scale of the true ground truth is much higher than that of the measurements -- as is the case with GONG. Likewise, we use the GONG PSF described in \cite{PlowmanEtal_2019I}. A number of test cases are considered.

\begin{figure}
		\begin{center}\includegraphics[width=\textwidth]{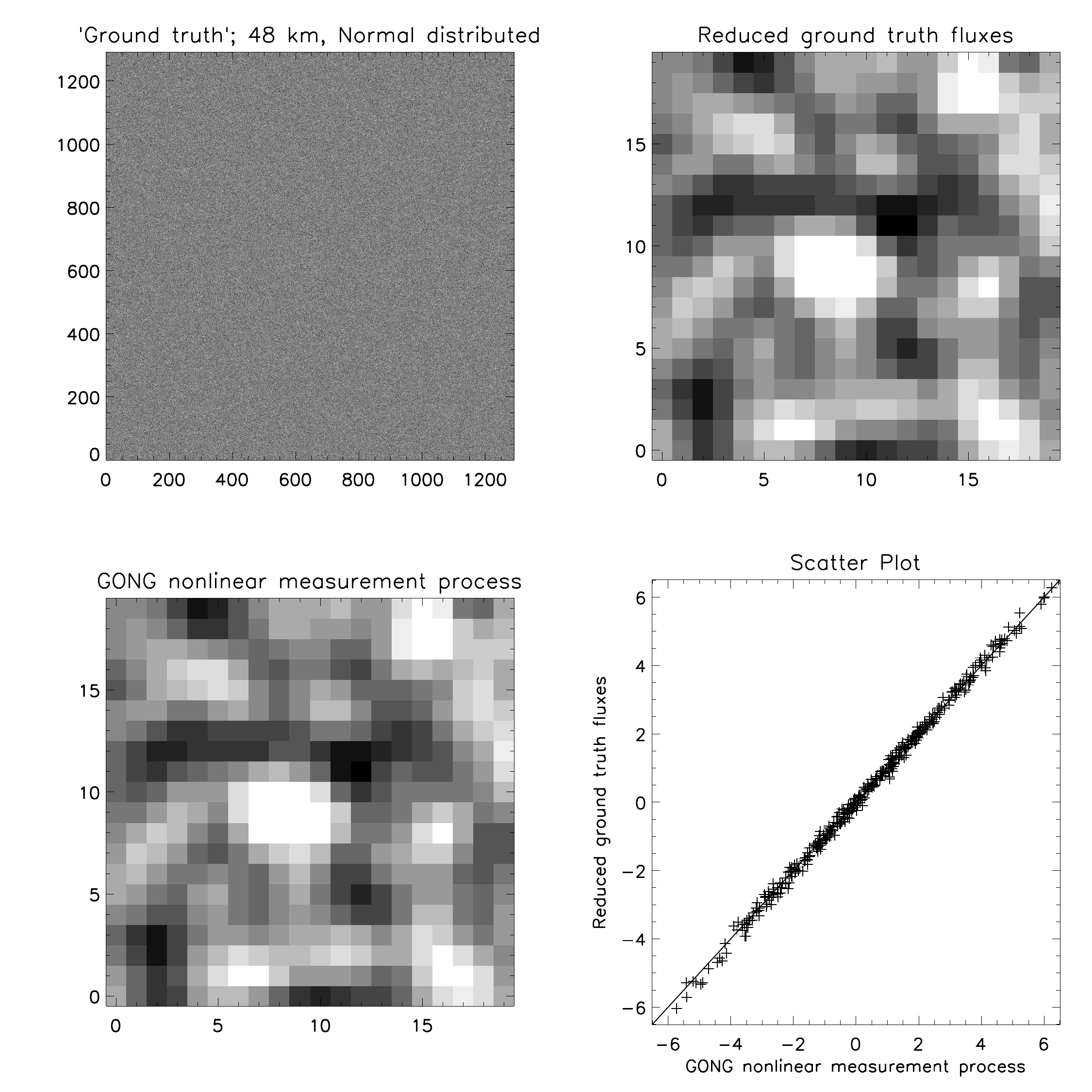}\end{center}
		\caption{Effects of GONG's plane-of-sky resolution combined with its nonlinear magnetic flux inversion, for uncorrelated Normal (Gaussian) distributed ground truth flux of standard deviation $500 G \cdot px$. The ground truth has $48\times 48$ km pixel size in this example. Top left shows the ground truth flux, top right shows the ground truth at GONG's resolution (i.e., convolved with PSF and pixelized), bottom left shows the GONG inversion, and bottom right shows the scatter plot between the inversion and the GONG-resolution ground truth. The nonlinear GONG inversion produces results that are essentially identical to the linear pixelization and PSF.}\label{fig:nonlinear_spatial_resolution_uncorrelated_hires}
	\end{figure}

	First is a random `salt and pepper' noise case, like that considered in previous sections -- the ground truth fluxes are uncorrelated and Gaussian distributed with mean 0 and standard deviation 500 Gauss. The spatial resolution of the ground truth was set to 48 km ($\sim 37$ ground truth pixels per GONG pixel). The results (Figure \ref{fig:nonlinear_spatial_resolution_uncorrelated_hires}) show no indication of significant effects resulting from the nonlinearity -- the slope of the ground truth vs. `measured' flux is unity to within 5\%, there is minimal scatter, and the image produced by the nonlinear GONG plane-of-sky sampling/inversion process is almost identical to that produced directly from applying pixelization and a linear PSF to the ground truth.

	We also checked whether the `native' spatial resolution of the ground truth affects the results in a similar `salt and pepper' noise case with spatial resolution 500 km ($\sim 3.5$ uncorrelated ground truth pixels per GONG pixel). The results (not shown) are essentially identical to the 48 km case, and again show no evidence of significant effects resulting from nonlinearity.

	To investigate whether stronger ground truth fluxes begin to show nonlinearity effects, we also checked another 48 km uncorrelated case, this time with a Normal distribution of standard deviation 1500 Gauss. This also showed a linear relationship with a slope of unity, although the scatter about that typical value was much higher. This further demonstrates that GONG is not susceptible to classical magnetograph saturation. There is an effect from outliers in the distribution wrapping around \citep[i.e., from $\sim 5$ kG to $-4.5$ kG; see][]{PlowmanEtal_2019I}, but this was checked and found to be small. An equivalent 500 km case (also not shown) had no obvious differences.

	Next we considered fluxes taken from the MURaM snapshot, rather than being uncorrelated random values, at a constant (photospheric) height slice of the simulation with inclination zero degrees. We found (also not shown) essentially the same result as the correlated and uncorrelated Normal distributed cases -- a linear relationship with minimal scatter, no offset, and near-unit slope. The 60 degree included MURaM case was also checked, with the same result. So far, so good.

\begin{figure}
		\begin{center}\includegraphics[width=\textwidth]{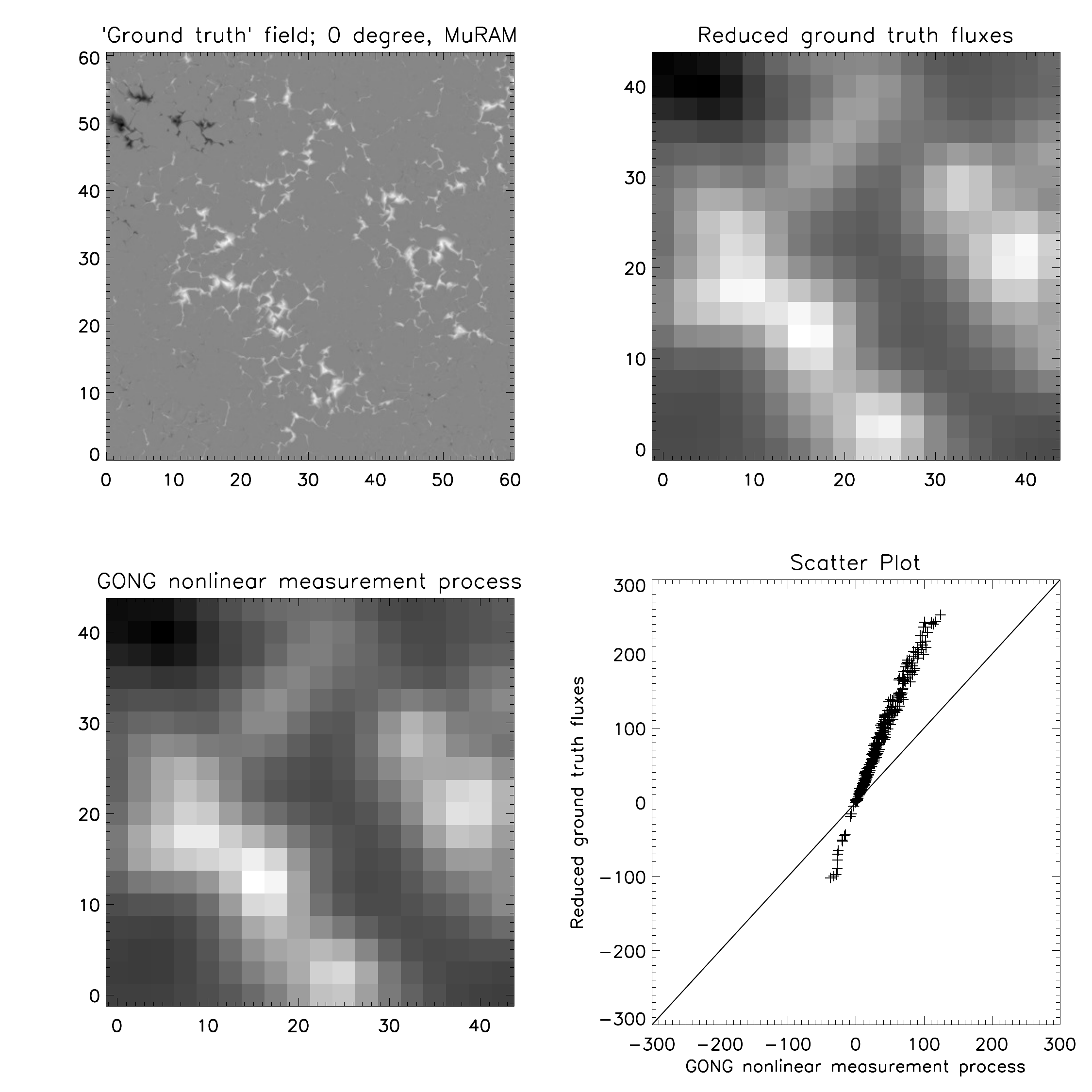}\end{center}
		\caption{Effects of GONG's plane-of-sky resolution combined with its nonlinear magnetic flux inversion, for a non-sunspot region of the MURaM snapshot at 0 degrees latitude. This now uses the Stokes $I$ profile computed by RH for each pixel, however Stokes $I\pm V$ still differ from Stokes $I$ only in the wavelength shift produced by the Zeeman effect. Top left shows the input (ground truth) MURaM flux, top right shows the ground truth at GONG's resolution (i.e., convolved with PSF and pixelized), bottom left shows the GONG inversion, and bottom right shows the scatter plot between the inversion and the GONG-resolution ground truth. Unlike the case with a constant Stokes $I$ profile (checked, but not shown), the nonlinear GONG inversion now produces results that underestimate the flux at small latitudes by over a factor of 2.}\label{fig:nonlinear_spatial_resolution_solar_0degrees_varprofile}
	\end{figure}

	However, these test cases all used the same reference Stokes $I$ wavelength profile for every pixel in the ground truth, and the Stokes $I\pm V$ profiles are the same as the Stokes $I$ profiles except that they are shifted in wavelength, in proportion to the flux -- the only difference between pixels is the strength of their fluxes, and therefore the size of the wavelength shift. The results change dramatically if we instead use the Stokes $I$ profile computed from the radiative transfer -- we simply substitute the Stokes I from the high resolution ground truth, with no other changes. Now the scatter plot for vertical viewing angle shows a slope of $\sim 2.4$ (Figure \ref{fig:nonlinear_spatial_resolution_solar_0degrees_varprofile}), instead of the slope of one found previously. Thus, it appears GONG will {\em underestimate} the flux by a factor of $\sim 2.4$ for these ground truth line profiles and fluxes. To investigate this, we repeated the experiment with the old reference profile, but changed its depth and line center to match that of the per-pixel profiles: a similar factor ($\sim 2.2$) was obtained. On the other hand, when the fluxes were set to uncorrelated Normal random values but the profiles from the radiative transfer were used, a slope of unity was obtained instead. 

	This implies that variations in the line depth and center, correlated with the locations of strong fields, can cause GONG to underestimate fluxes by a factor of 2 or more. This translates to an underestimation of net fluxes over larger scales, as we will show. Similar results were found when the resolution of the measurements was increased by a factor of 5 (nearly SDO/HMI resolution), while the effect was reduced  when the resolution was increased by a factor of 14 ($\sim 2$ times HMI). The effect is not restricted to GONG measurements: it is present to a lesser extent when the fluxes are computed from the low-resolution spectra using the center-of-gravity method \citep{Uitenbroek_ApJ2003}; there the factor is $\sim 1.5$.

	This effect appears similar to the `convective blueshift' \citep[see][for example]{Dravins_granulationspectra_AA1981,LohnerBottcher_ConvBlueshift_AA2018}. There, solar granulation appears to be blueshifted at lower resolutions because the granules and inter-granular lanes are unresolved: at disk center, the granule centers are brighter than the lanes and are blueshifted, whereas as the lanes are dark and redshifted. The inferred Doppler shift is weighted toward the brighter features (the centers), so it is blueshifted. A very similar explanation could be applied here: the bright granule centers have weak field while the dark lanes have strong fields, so the measured fluxes (inferred in a very similar way to the Doppler shift) will be weighted toward the weak values. 
	
\begin{figure}
		\begin{center}\includegraphics[width=\textwidth]{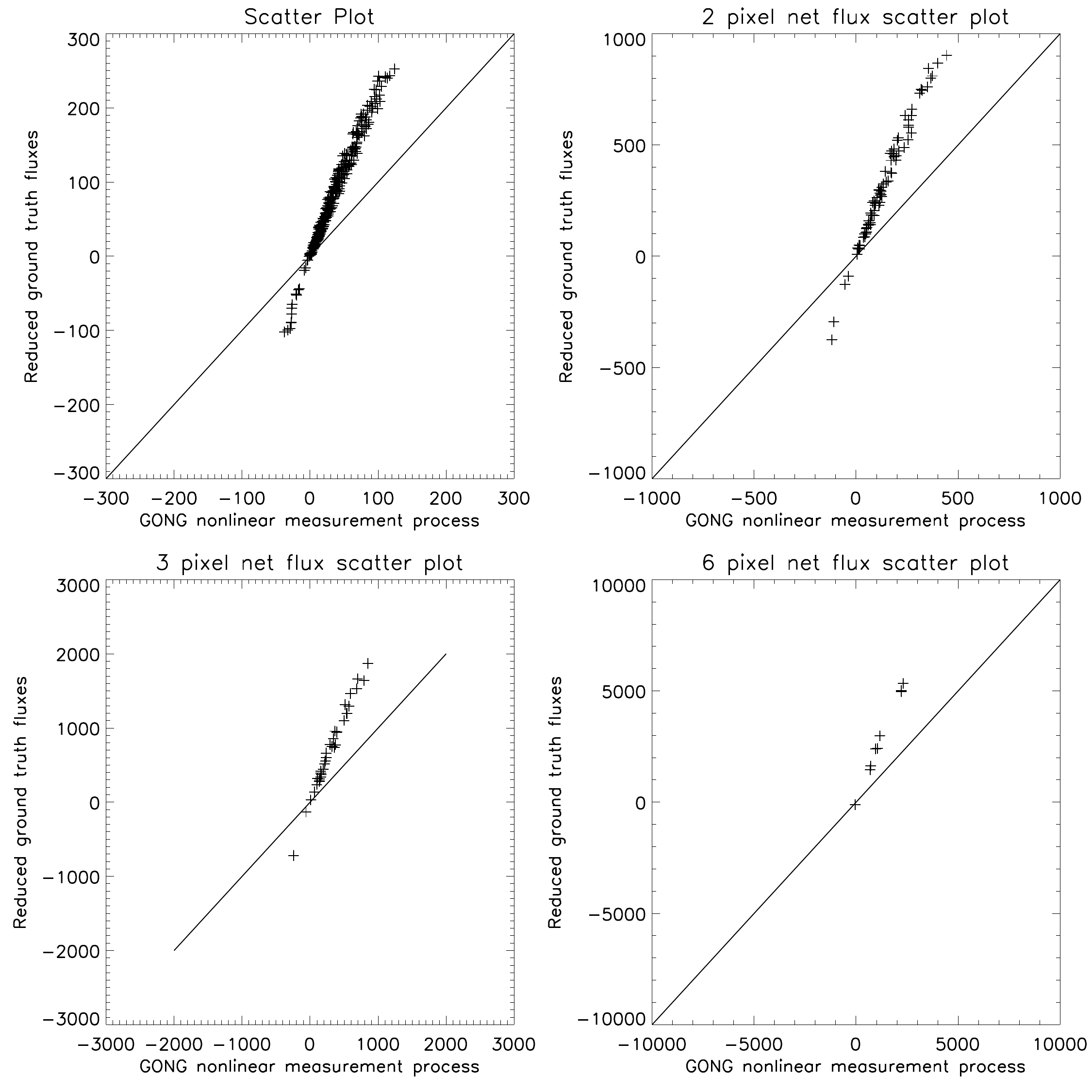}\end{center}
		\caption{Scatter plots showing net fluxes for the 0 degree MURaM case (Figure \ref{fig:nonlinear_spatial_resolution_solar_0degrees_varprofile}), with the ground truth placed at the same resolution as the measurements. The same scatter plot relationship is obtained for all net flux region sizes shown ($2\times 2$ pixels, $3\times 3$ pixels, and $6\times 6$ pixels), indicating that a calibration curve derived from the resolution-matched GONG-resolution point cloud (top left) will conserve the large-scale flux.}\label{fig:nonlinear_flux_conservation_MURaM_0degree}
	\end{figure}

\begin{figure}
		\begin{center}\includegraphics[width=\textwidth]{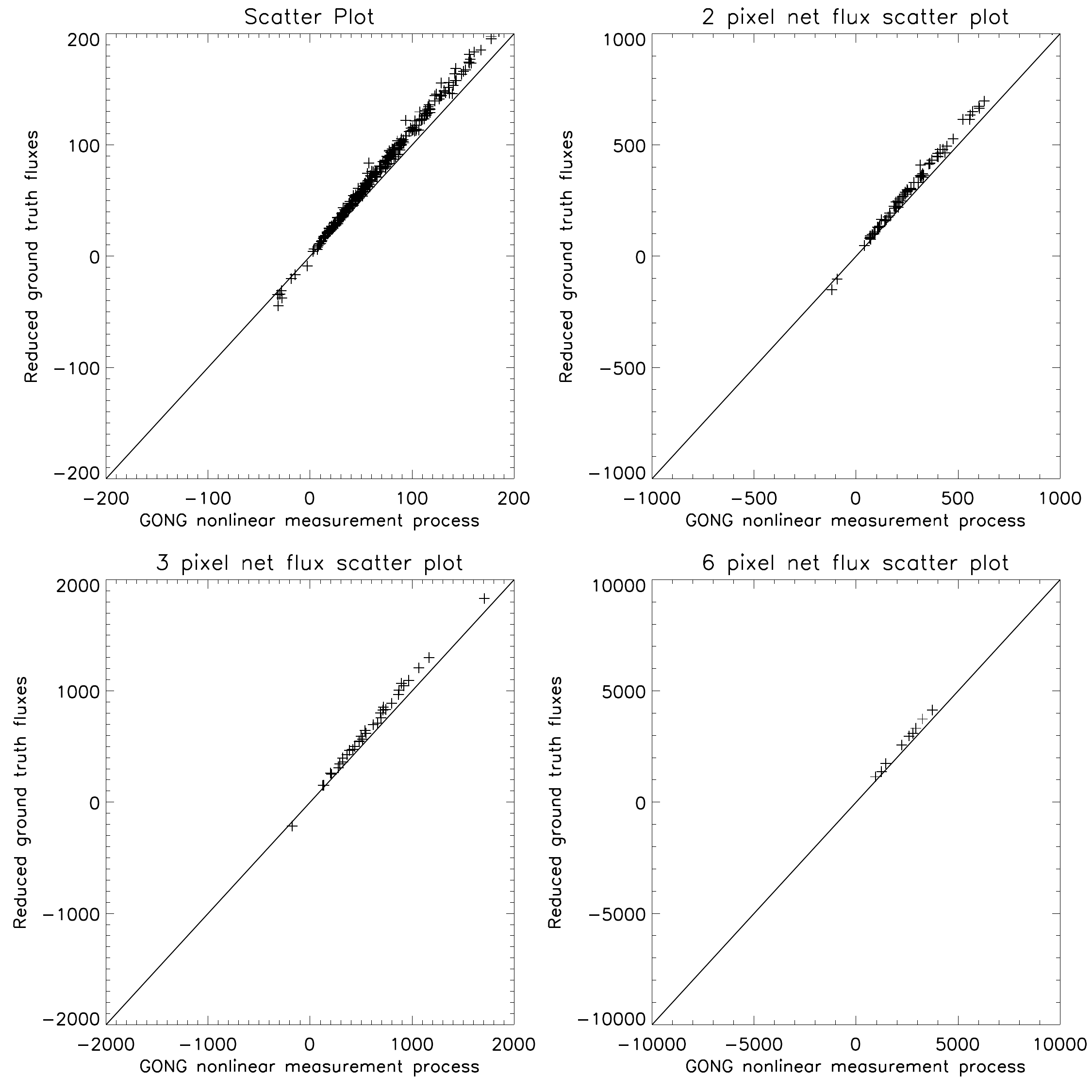}\end{center}
		\caption{Scatter plots showing net fluxes for the 60 degree MURaM case, with the ground truth placed at the same resolution as the measurements.  The same scatter plot relationship is obtained for all net flux region sizes shown ($2\times 2$ pixels, $3\times 3$ pixels, and $6\times 6$ pixels), indicating that a calibration curve derived from the resolution-matched GONG-resolution point cloud (top left) will conserve the large-scale flux.}\label{fig:nonlinear_flux_conservation_MURaM_60degree}
	\end{figure}

	In the case of the Doppler shift, the effect drops (and can even reverse) close to the limb because the contrast between the lanes and centers drops there (and can also even reverse). We checked in our GONG measurement test cases, and the underestimation factor does indeed drop. The top left panels of figures \ref{fig:nonlinear_flux_conservation_MURaM_0degree} and \ref{fig:nonlinear_flux_conservation_MURaM_60degree} clearly shown this for 0 and 60 degrees, and it will also be clearly demonstrated in the point clouds with the full simulator (the final paper in this series). The similarity of this behavior to the convective blueshift, and the similarity of the measurement process, is strong evidence that a similar effect is at play.

	Previous sections have demonstrated that the presence of a clear linear relationship in a scatter plot does not necessarily imply that a calibration curve derived from it will give the correct large-scale flux. To verify that, we need to see that the same clear linear relationship is maintained when the images used to make the scatter plot are rebinned to larger scales. Figure \ref{fig:nonlinear_flux_conservation_MURaM_0degree} shows that this is true for the vertical MURaM snapshot section, while Figure \ref{fig:nonlinear_flux_conservation_MURaM_60degree} shows that it is true for 60 degrees. In the final paper of this series, we will revisit this for all angles, with more points, and using the full GONG simulator.

	We have also verified that very similar point clouds to Section \ref{sec:linearmeasurement_solarexamples} are obtained in this nonlinear case when there is no resolution matching on the ground truth (i.e., only pixelization, no PSF), including the kink and offset at the origin. This reinforces the results of that section.
	
	\section{Summary \& Conclusions}\label{sec:groundtruthredux_sum}

	We have investigated the theory of magnetograph calibration, both in general and specifically to GONG, finding several significant conclusions in the process. First, we have pointed out that the most important aspect of magnetograph calibration is the preservation of flux: the calibrated magnetograms must not add, remove, amplify, or attenuate flux compared to the ground truth. They may very well spread it out, but that flux must come from somewhere on the magnetogram and go somewhere else on the magnetogram. The degradation of resolution caused by a PSF does exactly this, and even averaging over a pixel is explicitly a spreading out of flux compared to the high resolution ground truth.

	Moreover, for many of the space weather applications of magnetograms, and for GONG in particular, it is the flux at large scales ($\sim 0.1 R_{\odot}$) which is most important. Thus we conclude that removal of GONG's PSF (which is much smaller than $0.1 R_{\odot}$) is not the highest priority. We go on to point out that the available means of removing GONG's PSF are undesirable, either because they can introduce errors in the calibration (e.g., deconvolution) or because the level of effort is both impractical in the present work and not required given the goals of this work. We therefore specialize to `per-pixel' magnetogram comparison and calibration methods, which compare fluxes at each pixel between the two magnetograms independently of its neighbors.
	
	It has been suggested that, if a `ground truth magnetogram' with {\em no} PSF is compared with a `synthetic magnetogram' {\em with} PSF using a per-pixel method, a calibration made from that comparison would, in some `statistical' sense, `control' for the effect of the PSF. However, we have shown that this is not the case: Even in the simplest case (the PSF is applied by a linear convolution and normalized), the effect of the PSF is, by definition, a redistribution: It affects neighboring pixels in a correlated fashion (if one is diminished, its neighbors will be increased by the same amount). Per-pixel calibration, in contrast, treats each pixel in isolation: The correction for a given pixel depends only on its own flux value, with no reference to the fluxes of those around it. Therefore, a per-pixel calibration cannot reverse, control for, or even replicate the effects of a PSF.

	Although such per-pixel methods cannot capture or reverse the effects of a PSF, we have shown that they are still affected by it. If the ground truth is dominated by large flux variations between adjacent pixels, these will largely cancel as the PSF spreads the flux out: The PSF will reduce the amplitude of these variations, and therefore the `typical' {\em per-pixel} flux, even if it is flux-conserving (i.e., it does not add or remove flux from the magnetogram, for instance when it is applied by a linear convolution). A per-pixel comparison between magnetograms with differing PSFs (or between a ground truth without PSF and a synthetic measurement with one) will see this diminution in the magnitude of the per-pixel flux, and the scatter plots will show a non-unit slope even if the PSF is flux conserving. A calibration curve made from such a scatter plot will amplify the flux in the magnetogram and therefore violate flux preservation. Therefore, per-pixel comparison and calibration {\em cannot} be used to remove or `control' for the effects of a PSF, even in the simplest case and in a statistical sense.
	
	Fortunately, this issue can be avoided by matching the resolution of the ground truth to that of the synthetic measurements. We find that this produces matching fluxes in both the linear and the nonlinear case (Section \ref{sec:psfnonlinearity_examples}).  If the resolution of the measurements varies over the image plane (as it does in the real world with seeing), then the matching resolution on the ground truth must as well. However, our simulations employ a constant seeing size over the image plane, so this will not be necessary in our comparisons.

	We also find that `per-pixel' (e.g., point clouds or histogram equating) comparisons between two real magnetographs can be easily mislead by this same resolution mismatch, causing them to appear miscalibrated when they are not. A review of the literature (Appendix \ref{sec:resmismatch_literature}) finds that the PSFs are not always carefully considered when comparing, so a significant fraction of the apparent disagreement between contemporary magnetograms likely results from this. Like the ground truth comparisons, the resolution of the magnetograms must be carefully matched prior to making comparisons, including variation with time and over the image plane (due to seeing). If the PSFs are not well known, this can be accomplished by reducing both to much lower resolution. Only comparisons between magnetographs are subject to this effect, while extrapolations will not be. This would explain why \cite{Riley_comparison2014} finds a factor of $\sim 2$ difference between GONG and HMI, yet \cite{LinkerOpenFlux2017} finds that the extrapolated open flux for the two instruments are almost identical. Some publications comparing magnetograms \citep[e.g.,][]{2010LambEtal_ApJ720_140, 2012LiuEtal_SoPh279_295L, 2013PietarilaEtal_SoPh282_91} have also found a need to match resolutions, so this already has a peer reviewed track record; we supplement it by pointing out that it's absence can masquerade as a (spurious) calibration difference.
	
	We have chosen to apply the PSF (and atmospheric seeing effects) by a convolution in the spatial domain. This is entirely equivalent to Fourier filtering by the appropriate transfer function. For example, applying a low-pass filter in the Fourier domain is entirely equivalent to a convolution (by the fourier transform of the low-pass step function) in the spatial domain, by the convolution theorem.

	Finally, and most importantly, even when the resolution mismatch is accounted for, we find that the nonlinearity of the measurement process (Section \ref{sec:psfnonlinearity_examples}) results in a {\em genuine} underestimation of the fluxes, which is likely related to the `convective blueshift': For disk center viewing angles, unresolved variations in the line depth and center that are correlated with the locations of strong fields (e.g., the solar granulation pattern), can cause GONG to underestimate its flux measurements by a factor of 2 or more. This effect is related to instrument resolution, but only goes away when the resolution is high enough for the granulation pattern to be resolve: It is therefore likely that other synoptic instruments will be affected, but high resolution DKIST observations will not. The underestimation factor is much less at a 60 degree inclined viewing angle. We defer further discussion of this effect until \cite{PlowmanEtal_2019III}, where we peform the comparisons with the full GONG simulator results and investigate them in more detail.

	\subsection*{Acknowledgements}
	This work was funded in part by the NASA Heliophysics Space Weather Operations-to-Research program, grant number 80NSSC19K0005, and by a University of Colorado at Boulder Chancellor’s Office Grand Challenge grant for the Space Weather Technology, Research, and Education Center (SWx TREC).

	We acknowledge contributions, discussion, information, and insight from a variety of sources: Gordon Petrie, Jack Harvey, Valentin Martinez Pillet, Sanjay Gosain, and Frank Hill, among others.

	\section*{Disclosure of Potential Conflicts of Interest}
	The authors declare that they have no conflicts of interest.

	
	\begin{appendix}

	\section{Potential field source surface and flux conservation}\label{app:PFSS}
	
	Potential field extrapolations assume that the region of extrapolation is current-free and simply connected, in which case Maxwell's equations reduce to Laplace's equation, which can be solved by spherical harmonic expansion. The `source surface' assumption is that the field becomes radial at a fixed distance, $r_\mathrm{ss}$. This is motivated by the observation that, in white light, the `background' solar wind appears to stream radially outward at some radius, $r_\mathrm{ss}\approx 2.5 R_\odot$; the magnetic field outside this source surface is assumed to be dominated by the kinematics of the solar wind, resulting in currents which produce a radial field at $r_\mathrm{ss}$. The inner boundary is specified by the magnetogram, which leads to \citep[following][]{VirtanenMursula2017}
	\begin{equation}
		B_r(r,\theta,\phi) = \sum_{l=1}\sum_{m=0}^lP_l^m(\cos{\theta})C_l(r)[g_l^m\cos{(m\phi)}+h_l^m\sin{(m\phi)}]
	\end{equation}
	where $P_l^m(\cos{\theta})$ are the associated Legendre functions and (denoting the solar radius as $R_\odot$) the radial functions are
	\begin{equation}
		C_l(r)\equiv\Big[\frac{R_\odot}{r}\Big]^{n+2}\frac{n+1+n(r/r_\mathrm{ss})^{2n+1}}{n+1+n(R_\odot/r_\mathrm{ss})^{2n+1}}.
	\end{equation}
	The coefficients of the expansion, $g_l^m$ and $h_l^m$, are determined by integrating the solution against the magnetograms \citep[compare Equation 3 of][]{VirtanenMursula2017}:
	\begin{equation}\label{eq:gcoefficients}
		g_l^m = \frac{2n+1}{4\pi}\int_0^{2\pi}\int_{-1}^{1}\frac{B_\mathrm{LOS}(\theta,\phi)}{\sin{\theta}}P_l^m(\cos{\theta})\cos{(m\phi)}d\phi d\sin{\theta},\quad\mathrm{and}
	\end{equation}
	\begin{equation}\label{eq:hcoefficients}
		h_l^m = \frac{2n+1}{4\pi}\int_0^{2\pi}\int_{-1}^{1}\frac{B_\mathrm{LOS}(\theta,\phi)}{\sin{\theta}}P_l^m(\cos{\theta})\sin{(m\phi)}d\phi d\sin{\theta},
	\end{equation}
	where $B_\mathrm{LOS}(\theta,\phi)$ are the magnetogram values as functions of latitude and longitude and the $\sin{\theta}^{-1}$ is due to the assumption that the photospheric magnetic field is radial. As a result, the dependence of the radial field on the magnetogram is of the form
	\begin{equation}
		B_r(r,\theta,\phi) = \sum_i\int_0^{2\pi}\int_{-\pi/2}^{\pi/2}f_i(r,\theta,\phi,\theta',\phi')B_\mathrm{LOS}(\theta',\phi')d\phi d\theta,
	\end{equation}
	where $f_i$ rolls all of the terms into a single index, i. Thus, it is apparent that the dependence of the extrapolation on the magnetogram values is linear: If the calibration is a constant factor, its only effect on the extrapolated fields will be to multiply them by that same factor.  
	
\begin{figure}
		\begin{center}\includegraphics[width=\textwidth]{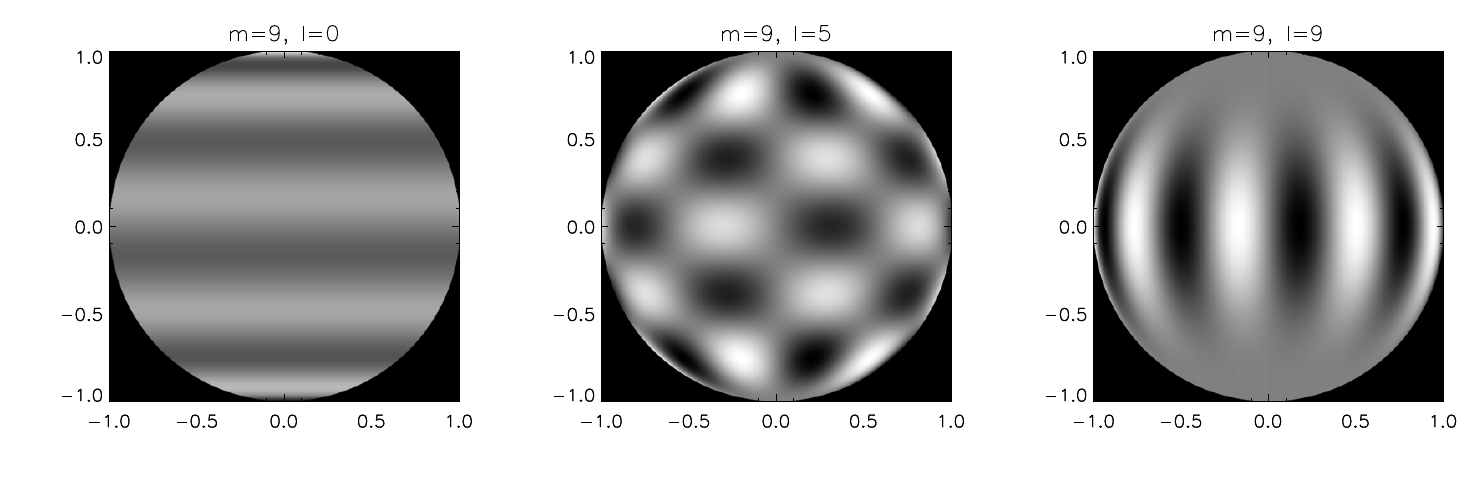}\end{center}
		\caption{Example $l=9$ spherical harmonics. Higher spatial frequencies than these do not contribute to PFSS extrapolations at the source surface.}\label{fig:sphericalharmonic_example}
	\end{figure}

	The harmonic expansion is only valid between $R_\mathrm{\odot}<r\leq r_\mathrm{ss}$, and it is the value of the PFSS at $r=r_\mathrm{ss}$ that is important for modeling: there, the radial functions are
	\begin{equation}
		C_l(r_\mathrm{ss}) = \frac{2l+1}{l+1+l(R_\odot/r_\mathrm{ss})^{2l+1}}\Big[\frac{R_\odot}{r_\mathrm{ss}}\Big]^{l+2}.
	\end{equation}
	This drops off rapidly with the order of the harmonic expansion: For $r_\mathrm{ss}=2.5 R_\odot$, the typical value, the $l=9$ term is smaller than the leading order ($l=1$) term by a factor of $\sim 750$, and each succeeding term is smaller than its predecessor by a factor of 2.5. Thus it is only the low order terms in the expansion which are important to these models -- this has been validated by \cite{KoskelaEtal2017}. Figure \ref{fig:sphericalharmonic_example} shows some of the $l=9$ spherical harmonics to illustrate their spatial scales.

	The upshot is that only large-scale, slowly-varying structure contributes to the field extrapolation at the source surface. The fields appear only as integrals against the slowly varying harmonic coefficients, i.e., as fluxes, so a calibration that preserves the large-scale flux will produce the same potential field extrapolation as the original ground truth. The order of the harmonic expansion sets the maximum spatial resolution required for the modeling; As long as the calibrated data, averaged to this resolution, matches the 'ground truth', averaged in the same way, the PFSS extrapolations will be in good agreement. In other words, the measurements must conserve flux at large scales, and any variations faster than the $l\sim 9$ spherical harmonics are less important for the extrapolations. If, on the other hand, these low-resolution averages do not match, either by some constant factor or with more complex variation, then a potential field extrapolated from the measurements will not match one made from the ground truth. 

	In particular, the spatial resolution of GONG is more than sufficient for the task, even with its PSF blurring. We have checked this by means of an experiment resampling HMI magnetograms to GONG plate scale. Two versions of the resampled magnetograms are produced: one which is simply resampled to GONG's pixel size (by bilinear interpolation to 5 times GONG resolution, then downsampling to GONG resolution with the IDL \texttt{rebin} function), while the other is convolved with the GONG's PSF and resampled (in the same way). We then computed a PFSS open flux map from each these magnetograms, using a similar method to \cite{Petrie2013}. There was no meaningful difference between them, demonstrating that the resolution difference introduced by the GONG PSF has no effect on PFSS open flux.

	\section{Curve fitting methods and resolution mismatch issues}\label{app:fittingmethods}
	
	In the ensuing derivations, we use $a$ (for actual) in place of $\phi^0$ (ground truths), $m$ for $\phi^m$ (measurements), $c$ for $\phi^c$ (calibrated measurements), and so on. This makes the math somewhat less cumbersome.
	
	\subsection{Forcing flux conservation by curve fitting?}\label{sec:gordonsfluxcurve}

	Suppose we attempt to overtly enforce flux conservation in the scatter plot curve fitting procedure, while still including the PSF on the $m_{ij}$ and omitting it from the $a_{ij}$. One suggestion has been looking for a curve where we split the  point cloud into bins in $m_{ij}$, and for each bin, set the calibration factor such that the net flux of the $c_{ij}$ in the bin is equal to the net flux of the $a_{ij}$ in the bin. That is, for each measured value bin, $m_l$, we take the calibration factor $\gamma_l$ to be
	\begin{equation}\label{eq:gordonsfluxcons_factor}
		\gamma_l = \Big(\sum_{m_l \leq m_{ij} < m_{l+1}} a_{ij}\Big)/\Big(\sum_{m_l \leq m_{ij} < m_{l+1}} m_{ij}\Big),
	\end{equation}
	where the summation is over all $i$ and $j$ such that $b_l \leq m_{ij} < b_{l+1}$. We then interpolate the $\gamma_l$ and $m_l$  to obtain the calibration factor for any given value of $m$,  $\gamma(m)$. The calibrated values are then
	\begin{equation}\label{eq:fluxcons_calvals}
		c_{ij} = m_{ij}\gamma(m_{ij})
	\end{equation}
	We begin, as before, with an analytic example. The point is to determine if there are issues with a proposed calibration scheme, and if issues are found with a simple test case, that is sufficient. The example is chosen for its simplicity and (relatively) easy analytic treatment, which has the advantage that the dependence of the calibration on its inputs can be seen algebraically. In this example, $a_{ij}$ are determined by probability density functions, in terms of which $\gamma_l$ can be written
	\begin{equation}
		\gamma_l = \frac{\int a p(a|m_l < m < m_{l+1}) da}{\int m p(m|m_l < m < m_{l+1}) dm}.
	\end{equation}
	These probabilities can be computed with the help of Bayes' theorem:
	\begin{equation}
		p(a|m_l < m < m_{l+1}) =  \frac{p(m_l < m < m_{l+1}|a)p_a(a)}{ p(m_l < m < m_{l+1})}
	\end{equation}
	\begin{equation}
		p(m|m_l < m < m_{l+1}) =  \frac{p(m_l < m < m_{l+1}|m)p_m(m)}{ p(m_l < m < m_{l+1})}
	\end{equation}
	\begin{equation}
		\gamma_l = \frac{\int a p(m_l < m < m_{l+1}|a)p_a(a) da}{\int p(m_l < m < m_{l+1}|m)p_m(m)dm} = \frac{\int a p(m_l < m < m_{l+1}|a)p_a(a) da}{\int_{m_l}^{m_{l+1}} p_m(m)dm}
	\end{equation}
	Here, $p_a(a)$ and $p_m(m)$ are the {\em marginal} probabilities of the `actual' and `measured' values, $a$ and $m$. We will once again assume a Normal probability density function for the $a_{ij}$:
	\begin{equation}\label{eq:normalgroundtruthpdf}
		p_a(a) = \frac{1}{2\pi\sigma_a^2}e^{-(a-a_0)^2/(2\sigma_a^2)}
	\end{equation}
	And as before, the $m_{ij}$ are related to the $a_{ij}$ by a scaling factor, $c_m$, a convolution. Now, we will also assume that the measurement errors are of the form $m_0+\Delta m_{ij}$, where $\Delta m_{ij}$ is Normal distributed with mean 0 and standard deviation $\sigma_{\Delta m}$:
	\begin{equation}
		m_{ij} = m_0+\Delta m_{ij} + c_m \sum_{kl}w_{ij,kl}a_{kl}
	\end{equation}
	Consider that each $m$ is the weighted sum (the weights are given by the kernel) of a set of random numbers with probability density function $p(a)$ plus $m_0+\Delta m_{ij}$. Since the sum of Normal random variables is itself a Normal random variable (with terms of the sum adding their standard deviations in quadrature), the errors in each of the $m_{ij}$ will also be Normal distributed, with mean 
	\begin{equation}\label{eq:linearmeasurementmean}
		\mu_m = a_0 c_m + m_0,
	\end{equation}
	and standard deviation equal to:
	\begin{equation}\label{eq:linearmeasurementsdev}
		\sigma_m =\sqrt{\sigma_{\Delta m}^2 + c_m^2\sigma_a^2\sum_{ij}K_{ij}^2} = \sqrt{\sigma_{\Delta m}^2 + c_m^2\frac{\sigma_a^2}{n_\mathrm{eff}}},
	\end{equation}
	where 
	\begin{equation}
		n_\mathrm{eff} \equiv 1/\sum_{ij}K_{ij}^2
	\end{equation}
	is the `effective' number of points in the kernel. If the kernel is a two-dimensional top-hat ($n_\mathrm{kernel}$ nonzero points, all equal to $1/n_\mathrm{kernel}$), for instance, then $n_\mathrm{eff} = n_\mathrm{kernel}$. If almost all of the weight comes from the central point ($1-K_{00}<<1$), then $n_\mathrm{eff}\approx 1$. In any case, $p_m(m)$ is then
	\begin{equation}\label{eq:linearmeasurementpdf}
		p_m(m) =\frac{1}{\sqrt{2\pi\sigma_m^2}} e^{-\frac{(m-\mu_m)^2}{2\sigma_m^2}}.
	\end{equation}
	The most challenging piece of these expressions is $p(m_l < m < m_{l+1}|a)$. To compute it, we must consider a modified version of $p_m(m)$, which we call $p_{m'}(m')$. This is the probability distribution of of $m_{ij}-a_{ij}$, that is, $m_{ij}$ computed with the central value of the kernel ($K_{00}$) zeroed out. This is also normally distributed, but because the new kernel is missing the $K_{00}$ term it only sums to $(1-K_{00})$. The mean is instead
	\begin{equation}
		\mu_m' = a_0c_m(1-K_{00})+ m_0,
	\end{equation}
	and the standard deviation is 
	\begin{equation}
		\sigma_{m'} = \sqrt{\sigma_{\Delta m}^2+\frac{c_m^2\sigma_a^2}{n_\mathrm{eff}'}},
	\end{equation}
	where $n_\mathrm{eff}'$ is defined similarly to $n_\mathrm{eff}$ (e.g., if the kernel is a two-dimensional top hat, then $n_\mathrm{eff}'=n_\mathrm{kernel}-1$):
	\begin{equation}
		n_\mathrm{eff}' = \frac{1}{\sum_{ij} K_{ij}^2-K_{00}^2} = \frac{n_\mathrm{eff}}{1-n_\mathrm{eff}K_{00}^2}
	\end{equation}
	The probability density function of $p_{m'}(m')$ is
	\begin{equation}
 		p_{m'}(m') = \frac{1}{\sqrt{2\pi\sigma_{m'}^2}}e^{-(m'-\mu_m')/(2\sigma_{m'}^2)}.
	\end{equation}
	Since this is the probability distribution of $m$ with the central pixel omitted from the convolution, it follows that the probability $m$ for a given central pixel value $a$ is the probability that $m'$ is equal to $m$ minus the central pixel's contribution, $ac_mK_{00}$\footnote{Note that we can't just compute that probability as $p(m+aK_{00})$ because $p(m)$ assumes a random value for the pixel, not the given value $a$.}:
	\begin{equation}
		p(m|a) = p_{m'}(m - ac_mK_{00})
	\end{equation}
	Then, 
	\begin{equation}
		p(m_l < m < m_{l+1}|a) = \int_{m_l}^{m_{l+1}}  p_{m'}(m - aK_{00})dm
	\end{equation}
	The calibration factors are then
	\begin{equation}
		\gamma_l = \frac{ \int a\int_{m_l}^{m_{l+1}}  p_{m'}(m - aK_{00})dm p(a) da}{\int_{m_l}^{m_{l+1}} m p(m) dm}
	\end{equation}
	If the bins are chosen to be small (e.g., the continuum limit) compared to the variation in $p_m(m)$ and $p_{m'}(m')$, then the $m$ integrands become their values at $m=m_l$ and the $dm$ ones with $m_{l+1}-m_l$ (the latter factors cancel):
	\begin{equation}
		\gamma_l = \frac{ \int a p_{m'}(m_l - aK_{00})p(a) da}{m_l p_m(m_l)}
	\end{equation}
	With the probability densities defined as above, evaluating this expression is laborious but straightforward. The result is
	\begin{equation}
		\gamma_l = \frac{K_{00}c_m}{\frac{c_m^2}{n_\mathrm{eff}}+\frac{\sigma_{\Delta m}^2}{\sigma_a^2}}+\frac{1}{m_l}(a_0-(a_0c_m+m_0)\frac{K_{00}c_m}{\frac{c_m^2}{n_\mathrm{eff}}+\frac{\sigma_{\Delta m}^2}{\sigma_a^2}}).
	\end{equation}
	The result is more clear when expressed in terms of the calibrated values, $c_{ij}$, as we can see by applying Equation \ref{eq:fluxcons_calvals} and rearranging:
	\begin{equation}\label{eq:gordonsfluxcons_cal}
		c_{ij} = (m_{ij}-m_0)\frac{K_{00}c_m}{\frac{c_m^2}{n_\mathrm{eff}}+\frac{\sigma_{\Delta m}^2}{\sigma_a^2}}+a_0\Big[1-c_m\frac{K_{00}c_m}{\frac{c_m^2}{n_\mathrm{eff}}+\frac{\sigma_{\Delta m}^2}{\sigma_a^2}}\Big].
	\end{equation}	

	We encounter here the same issue that we found earlier -- the calibration only `works' when the average flux of the region in question exactly matches the average flux (flux bias) of the ground truth ($\Phi'_\mathrm{net}/n'_\mathrm{tot}=a_0$). This is true even if the calibration data set has no flux bias (again, mirroring does not solve the problem): consider when $a_0=0$ and $m_0=0$ (no flux bias in `actual' values or observations); then,
	\begin{equation}
		c_{ij} = \gamma(m_{ij})m_{ij} =  m_{ij}\frac{K_{00}c_m}{\frac{c_m^2}{n_\mathrm{eff}}+\frac{\sigma_{\Delta m}^2}{\sigma_a^2}}.
	\end{equation}
	So the calibrated net fluxes  for some data set ($m_{ij}'$, actual values $a_{ij}'$) other than the one used to produce the calibration would be:
	\begin{equation}\label{eq:binratio_fluxfactor}
		\sum_{ij\ \mathrm{region}} c_{ij}' = \sum_{ij\ \mathrm{region}} m_{ij}'\frac{K_{00}c_m}{\frac{c_m^2}{n_\mathrm{eff}}+\frac{\sigma_{\Delta m}^2}{\sigma_a^2}} = \frac{K_{00}c_m}{\frac{c_m^2}{n_\mathrm{eff}}+\frac{\sigma_{\Delta m}^2}{\sigma_a^2}}\sum_{ij\  c_m\mathrm{region}}\sum_{kl} w_{ij,kl}a_{kl}'.
	\end{equation}
	As before, the normalization of the PSF implies that 
	\begin{equation}
		\sum_{ij\ \mathrm{region}}\sum_{kl} w_{ij,kl}a_{kl}' = \sum_{ij\ \mathrm{region}} a_{ij}' = \Phi_\mathrm{region}'
	\end{equation}
	modulo edge effects. Therefore, this time we have
	\begin{equation}
		\sum_{ij\ \mathrm{region}} c_{ij}' =\Phi_\mathrm{region}'\frac{K_{00}c_m}{\frac{c_m^2}{n_\mathrm{eff}}+\frac{\sigma_{\Delta m}^2}{\sigma_a^2}}
	\end{equation}
	for the case where there is no flux bias in the values used to produce the calibrations, which is correct only if $\Phi_\mathrm{region}=0$, or there is a lucky coincidence between the PSF, the measurement error, the `true' calibration factor ($c_m$), and the variation in the `actual' values used to produce the calibration. Remarkably, even if the kernel is a delta function ($K_{00}=1$, $n_\mathrm{eff}$=1), the correct region flux is obtained only if the measurement errors are much smaller than the variation in the ground truth field values ($\sigma_{\Delta m}^2<<\sigma_a^2$)!

	Returning to our previous example case -- no measurement errors, but the kernel is not a delta function (it is a Gaussian of width one pixel), while $c_m=1$ -- Equation \ref{eq:binratio_fluxfactor} reduces to just
	\begin{equation}
		c_{ij}' = m_{ij}'\frac{K_{00}}{\sum_{kl}K_{kl}^2},
	\end{equation}
	and the `calibrated' net flux would be
	\begin{equation}
		\sum_{ij\ \mathrm{region}} c_{ij}' =\Phi_\mathrm{region}'\frac{K_{00}}{\sum_{kl}K_{kl}^2}.
	\end{equation}
	The size of the factor $K_{00}/(\sum_{kl} K_{kl}^2)$ depends on the the shape of the PSF. For the 1-pixel width Gaussian used as example previously (e.g., \ref{fig:simpleexample_scatterplot}), that factor is 1.5, so this calibration method (using this calibration ground truth) will also lead to an overestimation of the flux by a factor of 1.5. We have implemented the method and find exactly the same theoretical predicted slope. For Gaussian PSFs much larger than a pixel, the slope is instead 2. There is one type of PSF that would result in a factor of one with this proposed calibration scheme -- a top hat -- however, such PSFs are not normally encountered in reality (GONG's certainly isn't one) and this is only true if the measurement errors are small ($\sigma_{\Delta m}^2 << \sigma_a^2$).

	Enforcing bin-wise flux conservation has not solved the resolution mismatch problem. It only ensures flux conservation for regions whose properties match those of the simulated flux values used to produce the calibration. As before, it does so by mixing terms between the correction factor and the offset, which means that the correction factors derived in this fashion cannot be relied upon -- in general, they are only correct if the PSF is a delta function and the measurement error is much smaller than the variation in the flux values. The idea of explicitly incorporating flux preservation into the curve fitting is not a bad one, however (we will return to it in the final paper when computing our GONG calibration curves): the issue is with the resolution mismatch and not the fitting procedure.

	\subsection{Comparing magnetograms by histogram equating}\label{sec:histogram_equating}
	
	In the interests of completeness, let us also consider another method of comparing two magnetograms, often used in the literature \citep[e.g.,][]{Riley_comparison2014, WenzlerEtal2004, JonesCeja2001}. It has the advantage that the magnetograms being compared do not need to be coregistered, resized to the same scale, or otherwise put on any kind of direct correspondence. However, we will show that this method has an {\em implicit} dependence on the relative resolutions of the instrument if the solar magnetic fields being observed have structure smaller than the resolution of either instrument.

	This comparison method is most clearly described in \cite{WenzlerEtal2004}, and we base our description on that work. In it, the two magnetograms to be compared are each split onto positive and negative halves. Then, each of these ranges is split into bins by their quantiles -- for example, the 0 to 0.01999\dots quantile (i.e., 0 to 1.999\dots percentile) of the positive flux in magnetogram a might be assigned to positive magnetogram bin 1, 0.02 to 0.039999\dots to bin 2, and so forth. There results 4 sets of bins, for a given magnetogram comparison:
	\begin{enumerate}
		\item Positive magnetogram a,
		\item Positive magnetogram b,
		\item Negative magnetogram a,
		\item Negative magnetogram b.
	\end{enumerate}
	
	The histogram equating value for each bin is then set to the mean of the magnetogram values within the bin for that set, positive and negative halves are combined, and the resulting two sets of values (one for magnetogram a, one for magnetogram b) are plotted against each other, bin for bin.

	Let's see what this procedure results in for the linear measurement model and Gaussian random `ground truth' example distribution discussed in Section \ref{sec:spatialresolution}. In that case, we are comparing `ground truth' $a_{ij}$, which have mean $a_0$ and standard deviation $\sigma_a$, with `measurements' $m_{ij}$ resulting from the linear measurement model, which will consequently have mean and standard deviation given by Equations \ref{eq:linearmeasurementmean} and \ref{eq:linearmeasurementsdev}, respectively. As discussed in Section \ref{sec:gordonsfluxcurve}, both are normal distributed according to equations \ref{eq:normalgroundtruthpdf} and \ref{eq:linearmeasurementpdf}. The locations of the bins ($a_i^+, a_i^-, m_i^+$, and $m_i^-$) are therefore defined according to Normal probability density functions ($p_a$ for the ground truth and $p_m$ for the measurements) and the quantile values ($q_i$):

	\begin{equation}
		\frac{\int_0^{a_i^+} p_a(a) da}{\int_0^\infty p_a da} = q_i = \frac{\int_0^{m_i^+} p_m(m) dm}{\int_0^\infty p_m dm}
	\end{equation}
	for the positive bins, and
	\begin{equation}
		\frac{\int_{a_i^-}^0 p_a(a) da}{\int_{-\infty}^0 p_a da} = q_i = \frac{\int_{a_m^-}^0 p_m(m) dm}{\int_{-\infty}^0 p_m dm}
	\end{equation}
	for negative bins. As implemented, the values used to compute the curves are set to the average of the magnetogram values within the bin. However, the bins are much smaller than the width of the distribution ($p_a$ and $p_m$) so these averages can be replaced with the bin locations ($a_i^+, a_i^-, m_i^+$, and $m_i^-$) for purposes of this example; bin widths are usually chosen to be $\sim 1$ percentile, which is much less than a standard deviation ($\sim 30$ percentiles).
	
	Thus, the bin averages can be found by simply inverting the quantile formulas above. These can be made more clear by expressing them in terms of error functions:

	\begin{equation}
		\frac{\erf{\left(\frac{a_i^+-a_0}{\sigma_a\sqrt{2}}\right)}-\erf{\left(\frac{a_0}{\sigma_a\sqrt{2}}\right)}}{1-\erf{\left(\frac{a_0}{\sigma_a\sqrt{2}}\right)}} = q_i = \frac{\erf{\left(\frac{m_i^+-\mu_m}{\sigma_m\sqrt{2}}\right)}-\erf{\left(\frac{\mu_m}{\sigma_m\sqrt{2}}\right)}}{1-\erf{\left(\frac{\mu_m}{\sigma_m\sqrt{2}}\right)}}
	\end{equation}

	And similarly for the negative bins. We can invert this equation using the inverse error function to express $a_i^+$ in terms of $m_i^+$ (or vice versa):

	\begin{equation}
		a_i^+ = a_0+\sigma_a\sqrt{2}\inverf{} \left[ 
			\erf{}\left(\frac{a_0}{\sigma_a\sqrt{2}}\right) +
			\frac{1-\erf{}\left(\frac{a_0}{\sigma_a\sqrt{2}}\right)}{1-\erf{}\left(\frac{\mu_m}{\sigma_m\sqrt{2}}\right)}\left(
				\erf{}\left(\frac{m_i^+-\mu_m}{\sigma_m\sqrt{2}}\right) - \erf{} \left( \frac{\mu_m}{\sigma_m\sqrt{2}} \right) 
			\right) 
		\right]
	\end{equation}
	
	This expression is not restricted to comparing a measurement to the specific ground truth from which it was obtained, since the derivation makes no assumption about any correspondence between the $a$ and $m$: it can be modified for comparison between any pair of measurements ($m$ and $m'$), provided the histograms of each can be reasonably characterized by a normal distribution, by simply replacing $a_i^+$ with $m_i^{'+}$, $a_0$ with $\mu_{m'}$, and $\sigma_a$ with $\sigma_m'$. 

	In the simplest non-trivial case comparing measurements with ground truth, $c_m=1$, $a_0=0$, $m_0=0$, and $\sigma_a^2/n_\mathrm{eff} \gg \sigma_{\Delta m}$ (again, see equations \ref{eq:normalgroundtruthpdf} and \ref{eq:linearmeasurementpdf}), this cumbersome expression is drastically simplified:

	\begin{equation}
		a_i^+ = m_i^+\sqrt{n_\mathrm{eff}}
	\end{equation}

	and comparison between two sets of measurements is similarly

	\begin{equation}
		m_i^{'+} = m_i^+\sqrt{\frac{n_\mathrm{eff}}{n_\mathrm{eff}'}}.
	\end{equation}
	This results in a similar slope to the other comparison methods previously described -- for the 1 pixel width Gaussian PSF, The slope ($\sqrt{n_\mathrm{eff}}$) is 1.56. We have verified that this analytical result is replicated when the same comparison is made numerically.

	\subsection{Resolution Mismatch in the Literature}\label{sec:resmismatch_literature}

	A variety of articles have compared magnetograms, and not all of them appear (from their text) to be fully cognizant of this resolution mismatch issue. With respect to the histogram equating method in general, \cite{WenzlerEtal2004} say

	\begin{quotation}
		`The basic underlying assumption of this method is that SPM and MDI magnetograms differ only in the scale of the magnetic field. By comparing the relative number of pixels (as opposed to absolute numbers) with a certain magnetic field the two data sets become directly comparable despite the different pixel size.'
	\end{quotation}
	Here it appears that the `scale of the magnetic field' refers not to the {\em spatial} scale of the field measurements, but rather to the flux scaling of the magnetographs going from solar flux to measured flux. In the language of previous sections of this paper, this is the magnetogram calibration factor $c_m$. The first statement is correct, but the second is not -- two magnetic flux data sets which differ in their pixel size (or PSF size) do not differ only in their flux scale (i.e., the flux scaling of the magnetic field measurements), because they are measuring fluxes integrated over different areas. This is true even if the fluxes are each divided by their areas.

    Histogram equating can be used to compare any two quantitative data sets, no matter how heterogeneous; that, in and of itself, does not make the data sets `directly comparable'. In this case, the results of the comparison depend on the pixel size and the resolution in general, and can show an {\em apparent} calibration difference even if the instruments are both perfectly calibrated (see Section \ref{sec:histogram_equating}). Although the method has no {\em explicit} dependence on the spatial resolution of the instruments, it does have an {\em implicit} one. 

	\cite{WenzlerEtal2004} also make comparisons with a more direct comparison method (their Section 3.2). However, although in this section they rebin the SPM data to match the pixel size of MDI, `to ensure that no bias due to the different pixel sizes enters the analysis', they do nothing to ensure that no bias due to the different {\em PSF sizes} enters the same analysis. And we have shown that there is indeed a bias due to different PSF sizes. Figure \ref{fig:example_pointcloud_MURaM} shows a similar curve to the one minute average curve of \cite{WenzlerEtal2004} Figure 6 (except that the axes are reversed). It therefore seems likely that at least some of the differences they find are due to the different resolutions (both pixel size and PSF size) of their instruments, not to intrinsic calibration differences of the instruemtns (i.e., different flux scaling).

	Similarly, when \cite{Riley_comparison2014} perform per-pixel comparisons between magnetographs, the higher resolution magnetograph is resampled to the pixel size of the lower resolution one, but the resolution differences due to the PSFs are not taken into account. They also perform histogram equating comparisons; the pixel and PSF size differences are not taken into account here either. Consequently, these results are also likely to be affected by the resolution mismatch issue.

	Some other means of comparison will not be affected by resolution difference issues. \cite{VirtanenMursula2017}, for example, directly compare the coefficients of the multipole expansions. Due to their central role in potential field extrapolations, those comparisons directly show where the calibration effects (including resolution) enter into the comparison and where they don't. They don't need to match resolutions in their comparison because they use an explicitly spatially aware method: the resolution differences only show up in the very high order terms, where they're genuine (and where they have very little effect on the extrapolation, as described in Appendix \ref{app:PFSS}). The drawback is that their results are much more complex than a handful of calibration curves.

	This is a likely explanation for why \cite{Riley_comparison2014} find highly variable correction factors (for HMI vs SOLIS, for instance), even though \cite{VirtanenMursula2017} find `The mutual scaling between SOLIS and HMI is very good, and one single overall coefficient of approximately 0.8 would be a reasonable choice for those data sets.': In the former case, the per-pixel resolution difference causes an apparent difference between the magnetograms across the board (because the comparison method is not spatially aware), while the latter is only affected by the resolution differences at those resolutions (because it is spatially aware).

	Other papers comparing magnetographs have found a need to degrade the resolution of one magnetograph beyond matching pixel sizes (this is another way of ensuring the resolutions match). For example, \cite{2013PietarilaEtal_SoPh282_91} found that, when comparing space-based magnetograms with those from SOLIS/VSM, they needed to spatially smooth the space-based magnetographs in order to counter the effects of bad seeing in the VSM magnetograms. Similarly, when \cite{2010LambEtal_ApJ720_140} compare SOHO MDI and Hinode-NFI magnetograms, they convolve both with a spatial Gaussian, reducing both of instruments to a common (lower) resolution. And when \cite{2012LiuEtal_SoPh279_295L} compare SoHO/MDI with SDO/HMI, they carefully reduce the HMI data to MDI's resolution, including the difference in PSF sizes (which they estimate); that comparison should therefore be unaffected by the resolution mismatch issue. So, while these effects have not been noticed by some in the literature, others have taken them into account. In this paper, we supplement this by clearly demonstrating the issue, why it arises, and how to correct for it. \cite{2010LambEtal_ApJ720_140, 2012LiuEtal_SoPh279_295L}, and \cite{2013PietarilaEtal_SoPh282_91} demonstrate that this means of correction already has a peer-reviewed track record.

	\end{appendix}
	\bibliographystyle{apj}
	\bibliography{apj-jour,GONG_simulatorII}
\end{document}